\documentclass[review]{elsarticle}

\usepackage[hidelinks]{hyperref}

\usepackage{amssymb}
\usepackage{amsmath}
\usepackage{amsthm}
\usepackage[algoruled,linesnumbered,vlined]{algorithm2e}
\usepackage{url}
\usepackage{dcolumn,booktabs}
\usepackage{subcaption}
\usepackage{pdflscape}
\usepackage{float}

\newcolumntype{d}[1]{D{.}{.}{#1}}
\newcommand\mc[1]{\multicolumn{1}{c}{#1}}

\journal{Computers \& Operations Research}

\biboptions{authoryear}

\begin{document}

\begin{frontmatter}

\title{Preprocessing and Cutting Planes with Conflict Graphs}

\author[decom,decsi]{Samuel Souza Brito\corref{cor1}}
\cortext[cor1]{Corresponding author.}
\ead{samuelbrito@ufop.edu.br}

\author[decom]{Haroldo Gambini Santos}
\ead{haroldo@ufop.edu.br}

\address[decom]{Departamento de Computa\c{c}\~{a}o, Universidade Federal de Ouro Preto, Brazil}
\address[decsi]{Departamento de Computa\c{c}\~{a}o e Sistemas, Universidade Federal de Ouro Preto, Brazil}

\begin{abstract}
This paper addresses the development of conflict graph-based algorithms and data structures into the COIN-OR Branch-and-Cut (CBC) solver, including: $(i)$ an efficient infrastructure for the construction and manipulation of conflict graphs; $(ii)$ a preprocessing routine based on a clique strengthening scheme that can both reduce the number of constraints and produce stronger formulations; $(iii)$ a clique cut separator capable of obtaining dual bounds at the root node LP relaxation that are $19.65\%$ stronger than those provided by the equivalent cut generator of a state-of-the-art commercial solver, $3.62$ times better than those attained by the clique cut separator of the GLPK solver and $4.22$ times stronger than the dual bounds obtained by the clique separation routine of the COIN-OR Cut Generation Library; and $(iv)$ an odd-cycle cut separator with a new lifting module to produce valid odd-wheel inequalities. The average gap closed by this new version of CBC was up to four times better than its previous version. Moreover, the number of mixed-integer programs solved by CBC in a time limit of three hours was increased by $23.53\%$.
\end{abstract}


\end{frontmatter}


\section{Introduction}

Over the last years, Mixed-Integer Linear Programming has proven to be a powerful technique for modeling and solving a wide variety of combinatorial optimization problems, most of them of practical interest. Some notable applications include telecommunication network design~\citep{Orlowski2010}, protein structure prediction~\citep{Xu2003} and production planning~\citep{Pochet2006}.

A Mixed-Integer Linear Program (MILP) deals with the minimization (or maximization) of a linear objective function subject to one or more linear constraints, where at least one of the decision variables can only assume integer values. It can be written as:

\begin{equation}
\tag{MILP}
	c^{*} = min \ \{c^{T}x \ | \ Ax \leq b, \ l \leq x \leq u, \ x \in \mathbb{R}^{n}, \ x_j \in \mathbb{Z} \ \forall j \in I\}
\label{eqn:milp}
\end{equation}

\noindent where $c \in \mathbb{R}^{n}$ represents the objective function coefficients, $A \in \mathbb{R}^{m \times n}$ is the constraint matrix and $b \in \mathbb{R}^{m}$ is the right-hand side (RHS) of the constraints. Vectors $l \in (\mathbb{R} \cup \{-\infty\})^{n}$ and $u \in (\mathbb{R} \cup \{+\infty\})^{n}$ are the lower and upper bounds for the decision variables, respectively. Furthermore, $N = \{1, ..., n\}$ is the index set of the decision variables $x$ and $I \subseteq N$ contains the indices of the variables that need to be integral in every feasible solution. A MILP where all of its decision variables are binary (i.e., $0 \leq x_j \leq 1, x_j \in \mathbb{Z}, \forall j \in N$) is also called Binary Program.

A feasible solution of a MILP is a vector in the set:

\begin{equation*}
	X = \{x \in \mathbb{R}^{n} \ | \ Ax \leq b, \ l \leq x \leq u, \ x_{j} \in \mathbb{Z} \ \forall j \in I\}
\end{equation*}

\noindent A solution $x^{*} \in X$ of a MILP is optimal if $c^{T}x^{*} = c^{*}$. A lower bound on the optimal solution value of a MILP can be obtained by solving its LP relaxation. The LP relaxation of a MILP is obtained when the integrality requirements are omitted:

\begin{equation}
\tag{LP}
	\check{c} = min \ \{c^{T}x \ | \ Ax \leq b, \ l \leq x \leq u, \ x \in \mathbb{R}^{n}\}
	\label{eqn:lp}
\end{equation}

\noindent This information is commonly used by the MILP solvers to explore promising nodes of the branch-and-bound search tree and to prove the optimality of a given solution.

Improvements in computer hardware and the development of several techniques such as preprocessing~\citep{Savelsbergh1994, Meszaros2003, Achterberg2016}, heuristics~\citep{Fischetti2005, Danna2005} and cutting planes~\citep{Hoffman1993, Rebennack2009} have contributed toward large-scale MILPs being solved effectively. Preprocessing and cutting planes are part of a mechanism called automatic reformulation~\citep{VanRoy1987}, which is a key component of modern MILP solvers. The works of \cite{Bixby2007}, \cite{Achterberg2013}, and more recently \cite{Achterberg2016} show that disabling these features in two state-of-the-art commercial solvers results in large performance degradation.

An implicit structure used by modern MILP solvers in preprocessing and cut separation routines is the Conflict Graph (CG)~\citep{Savelsbergh1994}. Such graphs represent the logical relations between binary variables. There is a vertex for each binary variable and its complement, with an edge between two vertices indicating that the variables involved cannot both be equal to one without violating the constraints.

In this paper, we present conflict graph-based algorithms and data structures that were developed and included in the COIN-OR Branch-and-Cut (CBC) solver\footnote{\url{https://github.com/coin-or/Cbc}}. CBC is one of the fastest open-source MILP solvers nowadays and it is also a fundamental component used by Mixed-Integer Nonlinear solvers, such as SHOT~\citep{Kronqvist2016}, Bonmin~\citep{Belotti2009} and Couenne~\citep{Bonami2008}.

Initially, we proposed and implemented a conflict graph infrastructure, characterized by the efficient construction and handling of such graphs. Our routine for building CGs is an improved version of the conflict detection algorithm presented by \cite{Achterberg2007}, which extracts conflicts from knapsack constraints. The basis for the improvement is a new step for detecting additional maximal cliques without increasing the computational complexity of the algorithm. We also developed optimized data structures that selectively store conflicts pairwise or grouped in cliques to handle dense CGs without incurring excessive memory usage. The sequence in which similar cliques are discovered is exploited to store them compactly.

After implementing the infrastructure for CGs, we developed a preprocessing routine and two cut separators. The preprocessing routine is based on the concept of clique merging proposed by \cite{Achterberg2016} and consists of extending set packing constraints by the inclusion of additional conflicting variables. A greedy algorithm uses the information from the conflict graph to augment the cliques formed by the set packing constraints. After executing the clique extension algorithm, all constraints that become dominated are removed. Computational results show that our routine was able to reduce the number of constraints and strengthen the initial dual bounds for a great number of instances.

The two conflict-based cut separators that we developed are responsible for separating clique and odd-cycle cuts. They replace the clique and odd-cycle cut separators used by CBC, which are provided by the COIN-OR Cut Generation Library (CGL)\footnote{\url{https://github.com/coin-or/Cgl}}. CGL is a library with a collection of cut generators that can be used with other COIN-OR packages.

Our clique cut separator was capable of obtaining dual bounds at the root node LP relaxation which are even stronger than the ones provided by a state-of-the-art commercial MILP solver. The improvements in the dual bounds obtained by including only odd-cycle cuts were relatively small. However, the inclusion of odd-cycle cuts did not present negative side-effects and the execution of the routine to separate these inequalities is computationally inexpensive, allowing its use in a cutting plane strategy.

The average gap closed by the new version of CBC was noticeably better than the previous version. For some instance sets, the average gap closed by this new version was up to four times better. Moreover, the time spent proving the optimality for the MILPs decreased and more instances were solved in restricted execution times.

The rest of this paper is organized as follows. Section~\ref{cgraph} formally explains the basic approach for constructing CGs and presents our conflict graph infrastructure. Section~\ref{preprocessing} details our conflict-based preprocessing routine while our clique and odd-cycle separators are introduced in Section~\ref{cuts}. In Section~\ref{experiments}, we document the results from extensive computational experiments, analyzing individually each one of the proposed routines as well as the performance of the resulting new version of CBC. Finally, in Section~\ref{conclusions}, we conclude and examine possible future research directions.

\section{Building Conflict Graphs}\label{cgraph}

A Conflict Graph (CG) is a structure that stores assignment pairs of binary variables which cannot occur in any feasible solution. It consists of an undirected graph with a set of vertices $V = \{x_j, \bar{x}_j \ : \ j = 1, ..., n\}$ and a set of edges $E \subset V^2$.

In such structure, vertex $x_j$ represents the assignment of the associated variable to one ($x_j = 1$), while vertex $\bar{x}_j$ corresponds to set the variable to zero ($x_j = 0$). Thus, the notation $\bar{x}_j$ is used to denote the binary complement of variable $x_j$ (i.e., $\bar{x}_j = 1 - x_j$). The assignment pairs represented by the edges in a CG are used to derive logical relations and, consequently, the edge inequalities provided in Table~\ref{tabLogicalRelations}.

\begin{table}[H]
\small
\caption{All possible logical relations between binary variables $x_j$ and $x_k$.}
\begin{center}
\begin{tabular}{ccrr}
\toprule
edge & logical relations & \multicolumn{2}{c}{edge inequality}\\
\midrule
$(x_j, x_k)$ & $x_j = 1 \implies x_k = 0$ & $x_j + x_k$ & $\leq 1$\\
 & $x_k = 1 \implies x_j = 0$ & & \\\hline
$(\bar{x}_j, \bar{x}_k)$ & $x_j = 0 \implies x_k = 1$ & $(1 - x_j) + (1 - x_k)$ & $\leq 1$ \\
 & $x_k = 0 \implies x_j = 1$ & & \\\hline
$(\bar{x}_j, x_k)$ & $x_j = 0 \implies x_k = 0$ & $(1 - x_j) + x_k$ & $\leq 1$ \\
 & $x_k = 1 \implies x_j = 1$ & & \\\hline
$(x_j, \bar{x}_k)$ & $x_j = 1 \implies x_k = 1$ & $x_j + (1 - x_k)$ & $\leq 1$ \\
 & $x_k = 0 \implies x_j = 0$ & & \\
\bottomrule
\end{tabular}
\end{center}
\label{tabLogicalRelations}
\end{table}

\begin{figure}[ht]
\begin{center}
	\includegraphics[scale=1.3]{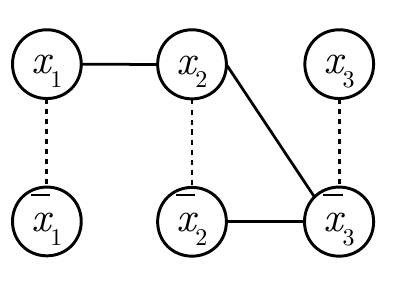}
	\caption{An example of a conflict graph.}
	\label{fig:cgraphex}
\end{center}
\end{figure}

An example of a conflict graph with three variables $\{x_1, x_2, x_3\}$ is presented in Figure~\ref{fig:cgraphex}. There is an edge linking each variable to its complement since only one must be equal to one in any feasible solution. Dashed lines in Figure~\ref{fig:cgraphex} denote these trivial conflicts. The conflict graph of this figure generates three edge inequalities:

\begin{alignat}{2}
	x_1 + x_2 \leq 1 & \nonumber\\
	x_2 + \bar{x}_3 \leq 1 &\Rightarrow x_2 - x_3 \leq 0\nonumber\\
	\bar{x}_2 + \bar{x}_3 \leq 1 &\Rightarrow x_2 + x_3 \geq 1 \nonumber
\end{alignat}

Generally, a graph with only a subset of the conflict edges is constructed, since discovering all pairs of conflicting assignments is \textit{NP}-hard. Deciding the feasibility of a binary program is \textit{NP}-complete~\citep{Garey1979}, and this task can be done by constructing and analyzing the full CG: a binary program is infeasible if and only if its associated CG is a complete graph.

CGs can be constructed using a probing technique based on feasibility considerations. This technique consists of tentatively setting binary variables to one of their bounds and checking whether the problem becomes infeasible as a result~\citep{Savelsbergh1994}. Thus, the edges of CGs can be obtained by analyzing the impact of fixing pairs of variables to different combinations of values.

The following subsection explains how the probing technique can be used to construct CGs, according to the work of \cite{Atamturk2000}. For ease of presentation and understanding, the remainder of this section only considers binary programs. Despite this, all of the techniques presented can be applied to any MILP containing binary variables.

\subsection{Probing Technique Based on Feasibility Conditions}

Suppose we are analyzing a constraint with the format

\begin{equation}
    \sum_{j \in B} a_{j}x_{j} \leq b,\label{knpConstraint}
\end{equation}

\noindent where $B$ is the index set of binary variables $x$ with non-zero coefficients in this constraint. Suppose also that we are investigating the impact of fixing two binary variables $x_{p}$ and $x_{q}$ to values $v_1$ and $v_2$, respectively. A valid lower bound for the left-hand side (LHS) of this constraint considering the assignments $x_{p} = v_1$ and $x_{q} = v_2$ is:

\begin{equation*}\label{lbLHS}
L^{x_p = v_1, x_q = v_2} = v_1 \cdot a_{p} + v_2 \cdot a_{q} + \sum_{j\in B^{-} \setminus \{p, q\}} a_{j}, 
\end{equation*}

\noindent where $B^{-}$ is the index set of variables with negative coefficients in the considered constraint. In this case, we consider the activation of the variables with negative coefficients to decrease the value of $L^{x_p = v_1, x_q = v_2}$ as much as possible. If $L^{x_p = v_1, x_q = v_2} > b$, then there is a conflict between the assignments of $x_{p}$ and $x_{q}$. Thus, we insert the corresponding edge in the graph.

This computation is performed for each pair of variables in each constraint in order to obtain a CG. Therefore, given a MILP with $m$ constraints and $n$ variables, its associated CG is constructed in $O(m n^2)$ steps. For this reason, probing may be computationally expensive for MILPs with a large number of variables and dense constraints. Nevertheless, for some constraint types, a large number of conflicts can be discovered quickly without having to conduct a pairwise inspection. For instance, in set packing and set partitioning constraints each variable has a conflict with all others, explicitly forming a clique. These constraints can be written as:

\begin{alignat}{2}
	\sum_{j \in C} x_j & \leq 1, x_j \in \{0, 1\} \ \forall j \in C \tag{set packing} \\
	\sum_{j \in C} x_j & = 1, x_j \in \{0, 1\} \ \forall j \in C \tag{set partitioning}
\end{alignat}

\noindent Set packing and set partitioning constraints often appear in MILPs to represent the choice of at most one (or exactly one) decision over a set of possibilities.

Considering the importance of CGs in solving MILP and aiming to accelerate the process of building these structures, \cite{Achterberg2007} developed a fast algorithm to extract cliques from the constraints. We improve this algorithm by inserting an additional step that detects a higher number of maximal cliques. Additionally, we design and implement data structures that selectively store conflicts pairwise or grouped in cliques to handle dense CGs without incurring excessive memory usage. Details of our conflict graph infrastructure are given in the following subsections.

\subsection{Fast Detection of Conflicts}

One way to accelerate the construction of CGs is detecting conflicts involving several variables simultaneously without using the pairwise inspection scheme. Following this idea, \cite{Achterberg2007} developed an algorithm that extracts cliques in less-structured constraints, that is, constraints that do not form a clique explicitly, by only traversing the constraint once. In addition to improve the process of building CGs, the early detection of cliques also allows for the more efficient storage of the conflicts since explicit pairwise conflict storage can prove impractical for dense graphs. Thus, one can make use of special data structures where large cliques are not stored as multiple edges, like the one proposed by \cite{Atamturk2000}. Alternatively, graph compression techniques such as GraphZIP~\citep{Rossi2018} could be employed to represent CGs succinctly.

We developed an improved version of the algorithm presented by \cite{Achterberg2007} to construct CGs. This algorithm exploits the fact that any linear constraint involving only binary variables can be rewritten as a knapsack constraint similar to (\ref{knpConstraint}), with $b > 0$ and $a_{j} > 0$ for each $j$ in the index set $B$ of binary variables $x$. Sometimes, transformations on the linear constraints are necessary to rewrite them in this format: for a variable $x_j$ with a negative coefficient $a_{j}$, we must consider the absolute value $|a_{j}|$, replace the variable by its complement $\bar{x}_j$ and update the RHS by adding $|a_{j}|$. For instance, the linear constraint $x_1 + x_2 - 2 x_3 \leq 0$ can be rewritten as $x_1 + x_2 + 2 \bar{x}_3 \leq 2$.

Algorithm~\ref{algClqDetection} presents our strategy to detect cliques on a given knapsack constraint. The first step is to sort the index set of variables $B$ in non-decreasing order of their coefficients. Next, we check if there are cliques in the constraint, by considering the activation of the two variables with the largest coefficients (line~\ref{beginChecking}). If this assignment does not violate the RHS of the constraint, we can ignore the possibility of the existence of conflicts and the algorithm finishes (line~\ref{endChecking}). Otherwise, we perform a binary search to find the smallest $k$ in $B$ such that $a_{j_{k}} + a_{j_{k + 1}} > b$ (line~\ref{cd1}). Once we found the value of $k$, a clique $C$ involving variables $\{x_{j_{k}}, x_{j_{k+1}}, ..., x_{j_{n}}\}$ is detected (line~\ref{cd2}). This clique is then stored in clique set $\mathcal{S}$ (line~\ref{cd3}) and the algorithm continues.

\begin{algorithm}[ht]
\small
\caption{Clique Detection}\label{algClqDetection}
\KwIn{Linear constraint $\sum_{j \in B} a_{j}x_{j} \leq b$.}
\KwOut{Set of cliques $\mathcal{S}$.}

Sort index set $B = \{j_{1}, ..., j_{n}\}$ by non-decreasing coefficient value $a_{j_{1}} \leq ... \leq a_{j_{n}}$\label{algSort}\;

\If{$a_{j_{n - 1}} + a_{j_{n}} \leq b$}
{
    \label{beginChecking}
    \Return $\emptyset$\;
    \label{endChecking}
}

$\mathcal{S} \gets \emptyset$\;
Find the smallest $k$ such that $a_{j_{k}} + a_{j_{k + 1}} > b$\label{cd1}\;
$C \gets \{x_{j_{k}}, ..., x_{j_{n}}\}$\label{cd2}\;
$\mathcal{S} \gets \mathcal{S} \cup \{C\}$\label{cd3}\;

\For{$o = k - 1$ downto $1$}
{\label{beginBS}
    Find the smallest $f$ such that $a_{j_{o}} + a_{j_{f}} > b$\label{newClq}\;

    \If{$f$ exists}
    {
        $A \gets \{x_{j_{o}}\} \cup \{x_{j_{f}}, ..., x_{j_{n}}\}$\;
	    $\mathcal{S} \gets \mathcal{S} \cup \{A\}$\;
    }
    \Else
    {\label{endBS1}
   		\textbf{break}\;\label{endBS2}
    }
}
\Return $\mathcal{S}$\;
\end{algorithm}

After finding $C$, the algorithm then attempts to detect additional maximal cliques. The strategy proposed by \cite{Achterberg2007} consists of iteratively trying to replace the variable with the smallest coefficient in clique $C$ by one of the variables outside $C$, maintaining the clique property. The disadvantage of this approach is that the additional cliques always differ in only one variable from the initial clique $C$. As such, cliques formed by a subset of variables of $C$ and a variable outside $C$ are not detected on the current constraint. This situation is solved using our new step for detecting additional maximal cliques, which occurs at lines~\ref{beginBS}~through~\ref{endBS2} of Algorithm~\ref{algClqDetection}. For each variable at position $o$ outside clique $C$, a binary search is performed to find the smallest $f$ such that the assignment pair $(x_{j_{o}} = 1, x_{j_{f}} = 1)$ violates the constraint. If $f$ exists, then an additional clique $A$ formed by variable $x_{j_{o}}$ and the subset $\{x_{j_{f}}, ..., x_{j_{n}}\}$ of $C$ is detected and stored. The algorithm stops when the binary search finds no results. The failure to find a position $f$ indicates that there are no additional cliques on the constraint since the coefficients are ordered.

Algorithm~\ref{algClqDetection} detects and stores cliques in $O(n \log n)$ steps on a given constraint with $n$ variables. In this algorithm, sorting a constraint (line~\ref{algSort}) is $O(n \log n)$, detecting an initial clique (line~\ref{cd1}) is $O(\log n)$, storing an initial clique (\ref{cd2}~to~\ref{cd3}) is $O(n)$, and detecting and storing additional cliques (lines~\ref{beginBS}~to~\ref{endBS2}) is $O(n \log n)$. Thus, a conflict graph for a MILP with $m$ constraints and $n$ variables is constructed in $O(m n \log n)$ steps, since we run Algorithm~\ref{algClqDetection} for each constraint.

It is important to note that detecting and storing additional cliques would spend $O(n^2)$ steps if we explicitly store all the contents of these cliques. With this approach, we would have to iterate over all elements of a detected clique to store it. Consequently, the worst-case complexity of Algorithm~\ref{algClqDetection} would be $O(n^2)$. However, any additional clique $A$ that can be found by this algorithm is always formed by a subset $C'$ of the first clique $C$ and one variable outside $C$. For this reason, we implemented a data structure that stores only $C$ completely. For each additional clique $A$, we store a tuple containing the variable outside $C$ and the first position of $C$ where the subset $C'$ starts. Therefore, storing an additional clique is $O(1)$ and, consequently, the loop that extracts additional cliques (lines~\ref{beginBS}~to~\ref{endBS2} of Algorithm~\ref{algClqDetection}) is $O(n \log n)$. Details about the data structures used to store conflict graphs are given in the next subsections.

The following example illustrates how our algorithm for detecting cliques on constraints works and compares the detected conflicts with the ones that could be found by the approach developed by \cite{Achterberg2007}.

\paragraph{\textbf{Example}}Consider linear constraints:

\begin{alignat}{2}
-3 x_1 + 4 x_2 - 5 x_3 + 6 x_4 + 7 x_5 + 8 x_6 & \leq 2\nonumber\\
x_1 + x_2 + x_3 & \geq 1\nonumber
\end{alignat}

\noindent where all variables are binary. The first step involves rewriting the constraints as knapsack constraints:

\begin{alignat}{2}
3 \bar{x}_1 + 4 x_2 + 5 \bar{x}_3 + 6 x_4 + 7 x_5 + 8 x_6 & \leq 10\label{eqExmpKnp1}\\
\bar{x}_1 + \bar{x}_2 + \bar{x}_3 & \leq 2\label{eqExmpKnp2}
\end{alignat}

\noindent Both constraints are already ordered according to their coefficients. We begin by analyzing constraint~(\ref{eqExmpKnp1}). First, we check for the existence of cliques in this constraint. When we activate the two variables with the largest coefficients ($x_5 = 1$ and $x_6 = 1$), we obtain $a_5 + a_6 = 7 + 8 = 15 > 10$. For this reason, we cannot discard the existence of cliques in this constraint. As such, we must now determine the smallest $k$ such that $a_{j_{k}} + a_{j_{k + 1}} > 10$. In this case, for $k = 3$ we have $a_3 + a_4 = 5 + 6 = 11$. Consequently, clique $C = \{\bar{x}_3, x_4, x_5, x_6\}$ is detected and stored. The next step consists of finding cliques involving variables $\bar{x}_1$ and $x_2$ outside $C$. For variable $x_2$, we perform a binary search that returns $f = 5$ since $a_2 + a_5 = 4 + 7 = 11 > 10$. Therefore, clique $\{x_2, x_5, x_6\}$ is detected. Finally, for $\bar{x}_1$ the binary search finds that $a_1 + a_6 = 3 + 8 = 11 > 10$, returning $f = 6$. Thus, clique $\{\bar{x}_1, x_6\}$ is also discovered. It is important to note that if we used the algorithm proposed by \cite{Achterberg2007}, cliques $\{x_2, x_5, x_6\}$ and $\{\bar{x}_1, x_6\}$ would not have been detected. As the last step, we analyze constraint~(\ref{eqExmpKnp2}). For this constraint we can discard the existence of cliques, since $a_2 + a_3 = 1 + 1 = 2 \leq 2$.

\subsection{Space Efficient Data Structure}

Data structures that efficiently store the cliques extracted from constraints are crucial in our algorithm. As mentioned before, explicitly storing all elements of all cliques extracted from a constraint increases the computational effort to construct CGs.

In Algorithm~\ref{algClqDetection}, after an initial clique $C$ is found, any additional clique $A$ is always a subset $C' \subset C$ plus a variable outside $C$. Moreover, given a clique $C = \{x_{j_{k}}, x_{j_{k+1}}, ..., x_{j_{n}}\}$, any subset of $C$ that composes an additional clique $A$ always has the form $C' = \{x_{j_{l}}, x_{j_{l+1}}, ..., x_{j_{n}}\}$, where $l > k$. Thus, an additional clique $A = \{x_{j_{o}}\} \cup C'$ can be represented by a tuple containing variable $x_{j_{o}}$ and the first variable $x_{j_{l}}$ of $C$ where subset $C'$ starts.

In our data structure, nodes $\{0, …, n-1\}$ represent the original variables while nodes $\{n, …, 2n-1\}$ represent the complements of the variables. We use three arrays to store the extracted cliques. A two-dimensional array, referred to here as \texttt{first}, stores the initial cliques extracted from the constraints. Thus, \texttt{first}$[c]$ stores the elements of the first clique extracted from the $c$-th constraint. The array entry \texttt{size}$[c]$ contains the size of the $c$-th clique of \texttt{first}. The last array, denoted as \texttt{addtl}, stores the additional cliques. In this array, a clique is represented by a tuple of the form $(x_o, c, l)$, which means that it is composed by variable $x_o$ and all variables at positions $l$ to \texttt{size}$[c]$ of the $c$-th clique of \texttt{first}. 

Additionally, auxiliary arrays are used to store, for each variable, the indices of \texttt{first} and \texttt{addtl} that contain cliques involving this variable. These structures are used to implement queries on the CG. The array entry \texttt{adjfirst}$[j]$ contains the indices of cliques stored in \texttt{first} that involve variable $x_j$. The number of cliques in \texttt{first} that contain $x_j$ is stored in \texttt{sizeaf}$[j]$. Arrays \texttt{adjaddtl} and \texttt{sizeaa} work in a similar way, but considering the cliques stored in \texttt{addtl}.

\begin{figure}[ht]
\begin{center}
    \includegraphics[scale=0.5]{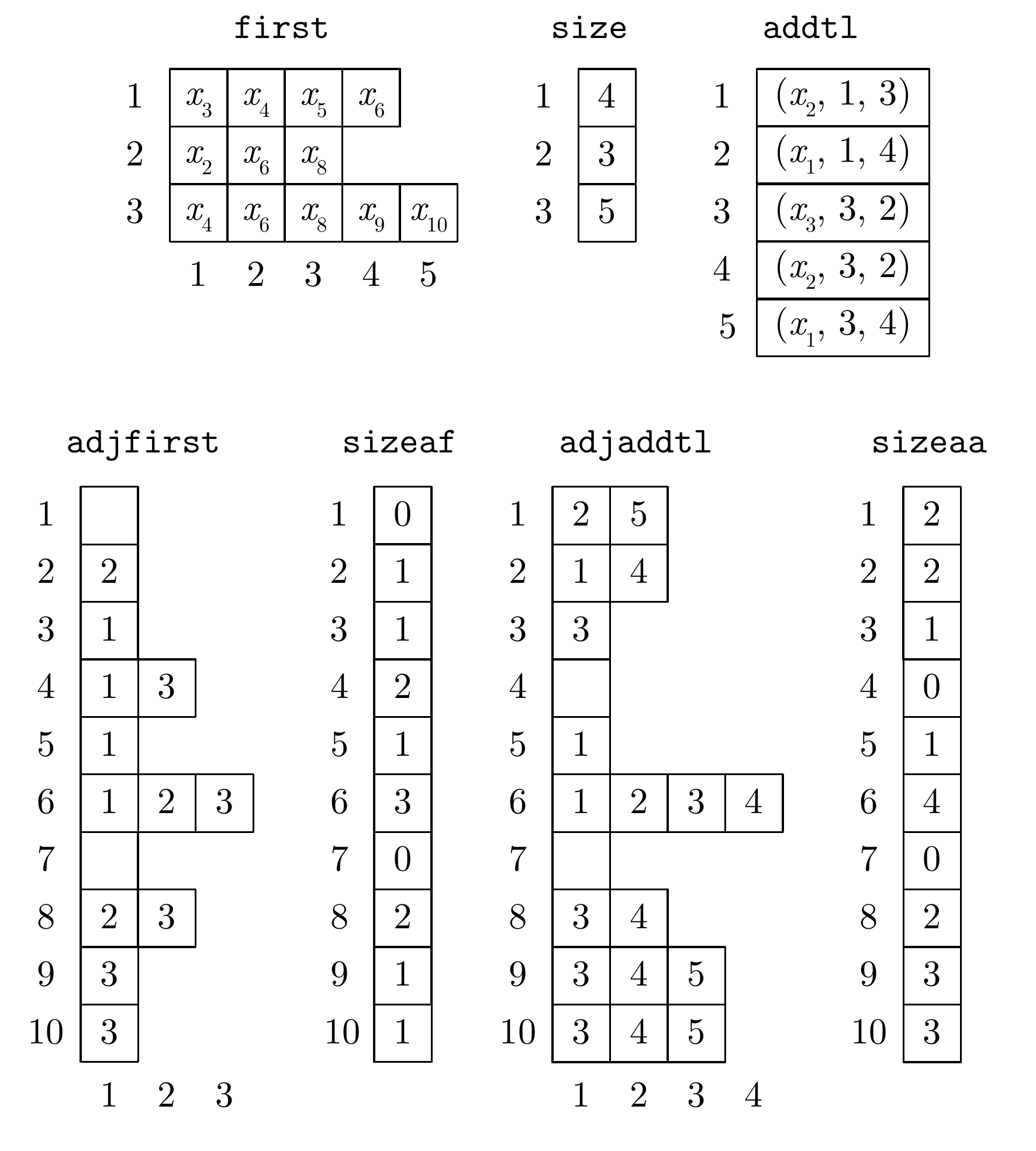}
    \caption{Filled data structure for the cliques of Table~\ref{tab:DataStr}.}
    \label{fig:datastr}
\end{center}
\end{figure}

Figure~\ref{fig:datastr} illustrates how our data structure works. This figure considers the cliques presented in Table~\ref{tab:DataStr}, which contains an example of cliques that can be found by Algorithm~\ref{algClqDetection}. We start storing the first clique $\{x_3, x_4, x_5, x_6\}$ of constraint $1$ at the first row of \texttt{first}. Then, the additional cliques of constraint $1$ are converted in tuples and inserted in \texttt{addtl}:

\begin{itemize}
    \item Clique $\{x_2, x_5, x_6\}$ is converted on tuple $(x_2, 1, 3)$, since it is composed by variable $x_2$ and subset $\{x_5, x_6\}$, which starts at index $3$ of the first clique of \texttt{first}.
    \item Clique $\{x_1, x_6\}$ is converted on tuple $(x_1, 1, 4)$, since it is composed by variable $x_1$ and subset $\{x_6\}$, which starts at index $4$ of the first clique of \texttt{first}.
\end{itemize}

\begin{table}[htbp]
\small
\caption{An example of cliques obtained after running Algorithm~\ref{algClqDetection} in three constraints.}\label{tab:DataStr}
\begin{center}
\begin{tabular}{cll}
\toprule
\mc{constraint} & \mc{first clique} & \mc{additional cliques}\\
\midrule
1 & $\{x_3, x_4, x_5, x_6\}$ & $\{x_2, x_5, x_6\}$ \\
 & & $\{x_1, x_6\}$ \\ \hline
2 & $\{x_2, x_6, x_8\}$ & \\ \hline
3 & $\{x_4, x_6, x_8, x_9, x_{10}\}$ & $\{x_3, x_6, x_8, x_9, x_{10}\}$ \\
 & & $\{x_2, x_6, x_8, x_9, x_{10}\}$ \\
 & & $\{x_1, x_9, x_{10}\}$ \\
\bottomrule
\end{tabular}
\end{center}
\end{table}

Following, clique $\{x_2, x_6, x_8\}$ of constraint $2$ is stored at the second row of \texttt{first}. This constraint does not have additional cliques. Finally, clique $\{x_4, x_6, x_8, x_9, x_{10}\}$ of constraint $3$ is stored at the third row of \texttt{first} and the three additional cliques of this constraint are stored in \texttt{addtl}:

\begin{itemize}
    \item $\{x_3, x_6, x_8, x_9, x_{10}\}$ is stored as $(x_3, 3, 2)$.
    \item $\{x_2, x_6, x_8, x_9, x_{10}\}$ is stored as $(x_2, 3, 2)$.
    \item $\{x_1, x_9, x_{10}\}$ is stored as $(x_1, 3, 4)$.
\end{itemize}

\subsection{Query Efficient Data Structure}

The use of the data structure previously presented allows reducing the computational effort required to build CGs. It not only accelerates the construction process but also decreases the memory required to store a graph. However, the cost of making queries in these data structures increases according to the number of cliques stored by it. For example, the worst case of a query that returns all variables conflicting with a given variable occurs when we iterate over all cliques of a graph. The execution of this query in a graph with a large number of cliques could take considerable time. This query would be faster if we use adjacency lists for each variable, explicitly storing all conflicting variables. In contrast, the use of adjacency lists increases the memory consumption and the computational effort required to build CGs, especially for the dense ones. Hence, there is a tradeoff between the computational effort and memory requirements to build and store a conflict graph and the performance of querying it.

We implemented a hybrid solution that tries to limit the memory consumption of the conflict graph without significantly affecting the time spent to construct and query this structure. This solution uses the data structure presented before and maintains an adjacency list for each vertex. The array of adjacency lists are referred to here as \texttt{adjlist} and each array entry \texttt{adjlist}$[j]$ contains a set of variables conflicting with variable $x_j$. The adjacency list of each vertex is kept sorted so that queries in it can be performed in $O(\log n)$. A parameter $minClqSize$ controls how the cliques are stored. After creating a conflict graph, we iterate over the cliques and remove those whose sizes are less than or equal to $minClqSize$. These small cliques are now stored as multiple pairs of conflicts in the adjacency lists of the vertices involved. Thus, large cliques are explicitly stored in \texttt{first} and \texttt{addtl}, while the small cliques are stored as multiple pairs of conflicts in \texttt{adjlist}.

Checking if two variables are conflicting is a query that frequently appears in conflict graph-based routines. Considering our hybrid solution, two variables $x_j$ and $x_k$ are conflicting if $x_k$ appears in the adjacency list of $x_j$ (or vice-versa) or if they appear together in at least one clique. Thus, the first step of this query method is to perform a binary search to test if $x_k$ exists in \texttt{adjlist}$[j]$. If \texttt{adjlist}$[j]$ contains $x_k$, variables $x_j$ and $x_k$ are conflicting and the method finishes. Otherwise, the method iterates over the cliques stored in \texttt{first} and \texttt{addtl} that contain $x_j$. For a given iteration, the method finishes if the current clique contains variable $x_k$, indicating that $x_j$ and $x_k$ are conflicting. After iterating over all cliques that contain $x_j$ and finding none of them that also contains $x_k$, the method concludes that $x_j$ and $x_k$ are not conflicting.

Another query method that is frequently employed in routines based on CGs is the one that returns all variables conflicting with a given variable $x_j$. Considering our data structures, this method initially creates a set $Q$ containing all variables of \texttt{adjlist}$[j]$. Then, it iterates over the cliques in \texttt{first} and \texttt{addtl} that contain $x_j$, storing in $Q$ all variables that are part of these cliques. Finally, set $Q$ is returned.

The queries in our conflict graph infrastructure are efficient when most of the cliques are stored as multiple pairs of conflicts in the adjacency list of the vertices. In fact, for typical instance problems of MIPLIB~\citep{Miplib2017}, the queries in the conflict graphs are faster when we set $minClqSize = 512$. In these conflict graphs, just a small set of conflicts are explicitly stored as cliques.

\section{Conflict-Based Preprocessing: Clique Strengthening}\label{preprocessing}

Preprocessing is an essential component in MILP solvers that can modify the structure of a MILP to produce a stronger formulation. Stronger formulations usually have tighter dual bounds, which makes the branch-and-bound process more effective. Thus, a preprocessing component may accelerate the solution process and enable the early detection of infeasible problems.

There are several preprocessing strategies proposed in the literature. One of the precursors of these strategies was the work of \cite{Brearley1975}, which describes techniques for mathematical programming systems that reduce the problem dimension by fixing variables, removing redundant rows, replacing constraints by simple bounds and more. \cite{Savelsbergh1994} presented a framework for describing preprocessing and probing techniques, providing an overview of simple and advanced techniques to improve the representation of MILPs. More recently, \cite{Gamrath2015} developed three preprocessing techniques that were included in the non-commercial solver SCIP, and \cite{Achterberg2016} described the preprocessing strategies implemented in the commercial solver Gurobi.

One of the preprocessing strategies developed by \cite{Achterberg2016} and included in Gurobi is called \emph{clique merging}. It consists of combining several set packing constraints into a single inequality. We based on this algorithm to develop a preprocessing routine that extends set packing constraints instead of combining them. We consider the whole conflict graph to extend each one of these constraints. Thus, variables that do not appear in other set packing constraints can be included in an extended constraint.

First, we create a set $\mathcal{C}$ containing all cliques formed by the set packing constraints of a given MILP. Then, we try to extend each clique $C$ in $\mathcal{C}$ using Algorithm~\ref{algClqExtension}.

\begin{algorithm}[ht]
\small
\caption{Clique Extension}\label{algClqExtension}
\KwIn{Conflict graph $G = (V, E)$ and clique $C$.}
\KwOut{Extended clique $C'$.}

Let $d$ be the vertex in $C$ with the smallest degree\label{candListB}\;
$L \gets \{k \in N_{G}(d) \ | \ k \notin C\}$\label{candListE}\;
$C' \gets C$\;
\While{$L \neq \emptyset$} {
	Let $l$ be the vertex in $L$ with the largest degree in $G$\;
	Remove $l$ from $L$\;

	\If{$\forall k \in C' : \ l \in N_G(k)$} {
	\label{tryInsertB}
		$C' \gets C' \cup \{l\}$\label{tryInsertE}\;
	}
}
\Return $C'$\;
\end{algorithm}

Algorithm~\ref{algClqExtension} is based on a greedy strategy that uses the information from a conflict graph $G = (V, E)$ to add variables in clique $C$. Initially, a set $L$ of candidate vertices for inclusion in clique $C$ is constructed by selecting all neighbors of a vertex $d$ that are not contained in $C$. Vertex $d$ is the one with the smallest degree between the vertices of $C$. Notation $N_{G}(d)$ is used to represent the vertices in graph $G$ that are adjacent to vertex $d$. Next, we create a set $C'$ that initially contains all vertices of $C$. Then, we try to insert additional vertices in $C'$ by iteratively selecting the vertex $l \in L$ with the largest degree in $G$. Vertex $l$ is inserted in $C'$ only if it is adjacent to all vertices in $C'$ (lines \ref{tryInsertB} and \ref{tryInsertE}). The algorithm finishes when $L$ is empty, returning the extended clique $C'$.

After obtaining clique $C'$, we generate the corresponding set packing constraint and insert it into the MILP. Finally, a dominance checking procedure is performed to remove all constraints that are dominated by this extended constraint~\citep{Achterberg2016}. In this context, a constraint $i'$ dominates another constraint $i$ if the corresponding clique of $i$ is a subset of the clique formed by $i'$.

Our clique strengthening routine is specially effective when applied to MILPs that have several constraints expressed by pairs of conflicting variables. However, it can be computationally expensive, specially for problems with dense CGs and constraints with a large number of variables. For this reason, we limit the execution of the preprocessing routine to constraints with at most $\alpha_{max}$ variables, where $\alpha_{max}$ is an input parameter of the algorithm.

The following example illustrates the execution of the clique strengthening process.

\paragraph{\textbf{Example}}Consider the following linear constraints

\begin{alignat}{2}
-4x_{1} + 4x_{2} + 5x_{3} + 6x_{4} + 7x_{5} + 10x_{6} &\leq 6 \label{clqS1}\\
x_{2} + x_{3} + x_{4} &\leq 1 \label{clqS2}\\
x_{2} + x_{5} &\leq 1 \label{clqS3}
\end{alignat}

\noindent The first step is to rewrite constraint~(\ref{clqS1}) as a knapsack constraint:

\begin{equation*}
    4\bar{x}_{1} + 4x_{2} + 5x_{3} + 6x_{4} + 7x_{5} + 10x_{6} \leq 10
\end{equation*}

\begin{figure}[ht]
    \begin{center}
        \includegraphics[width=0.4\textwidth]{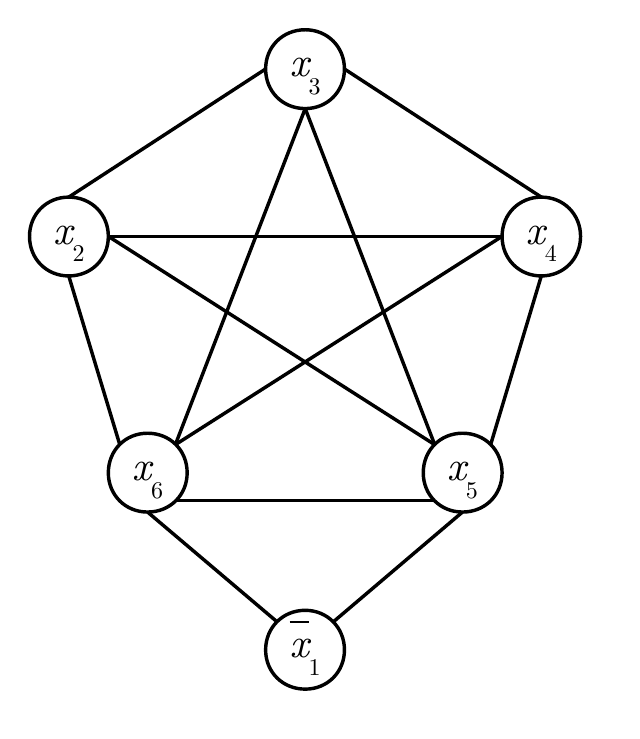}
        \caption{Conflict graph for constraints~(\ref{clqS1}) to (\ref{clqS3}).}\label{figExmpClqS}
    \end{center}
\end{figure}

\noindent Now, all the constraints are in the knapsack constraint format and we can run our algorithm for building the CG. Figure~\ref{figExmpClqS} shows the graph associated with constraints~(\ref{clqS1}) to (\ref{clqS3}).

\noindent Set $\mathcal{C}$ is then created, containing cliques of the constraints~(\ref{clqS2}) and (\ref{clqS3}). The clique strengthening procedure is first applied to constraint~(\ref{clqS2}), producing the extended constraint:

\begin{equation*}
    x_{2} + x_{3} + x_{4} + x_{5} + x_{6} \leq 1
\end{equation*}

\noindent Then, we remove constraints~(\ref{clqS2}) and (\ref{clqS3}) since they are dominated by this extended constraint. There are no more constraints in $\mathcal{C}$ to be extended. Thus, the execution of clique strengthening in constraints~(\ref{clqS1}) to (\ref{clqS3}) results in the following constraints:

\begin{alignat}{2}
-4x_{1} + 4x_{2} + 5x_{3} + 6x_{4} + 7x_{5} + 10x_{6} &\leq 6\nonumber\\
x_{2} + x_{3} + x_{4} + x_{5} + x_{6} &\leq 1\nonumber
\end{alignat}

\section{Cutting Planes}\label{cuts}

A primary application for conflict graphs is the generation of valid inequalities, also known as \emph{cuts}. Cuts are linear constraints $a^{T}x\leq b$ that are violated by the current LP solution of a MILP but do not remove any integer feasible solution. The addition of such inequalities enables tightening the LP relaxation of a MILP, approximating it to the convex hull of integer feasible points. Cutting planes are often combined with a branch-and-bound scheme, resulting in the branch-and-cut or cut-and-branch algorithms that are present in modern MILP solvers.

Any feasible solution of a MILP defines a vertex packing in its associated conflict graph. A vertex packing in a graph $G = (V, E)$ is a subset $P \subseteq V$ for which all $v_j, v_k \in P$ satisfy $(v_j, v_k) \notin E$. Based on this concept, one can conclude that any valid inequality for the vertex packing polytope is also valid for the convex hull of feasible solutions for this MILP~\citep{Atamturk2000}. Thus, conflict graphs can be used to find inequalities that cut off the current LP solution.

Cliques and odd cycles are some of the most common classes of inequalities derived from the vertex packing polytope. Generally, the improvement in the value of the LP relaxation obtained by the inclusion of odd-cycle inequalities is small~\citep{Borndorfer1998, Mendez-Diaz2008}. However, the execution of a routine to separate these inequalities is computationally inexpensive in comparison with other cut separators, since they can be separated in polynomial time using shortest path algorithms~\citep{Grotschel1993, Rebennack2009}.

The following subsections present our routines for separating these conflict-based cuts. Preliminary versions of our cut separators were successfully applied to three classical combinatorial optimization problems: Capacitated Vehicle Routing~\citep{Pecin2017}, Project Scheduling~\citep{Araujo2019} and Nurse Rostering~\citep{Santos2016}. The use of these routines contributed to the solution of several hard instances for the first time in the literature.

\subsection{Clique Cut Separator}

A clique inequality for a set $C$ of conflicting variables is defined as:

\begin{equation*}
    \sum_{j \in C} x_j \leq 1
\end{equation*}

\noindent where $C$ is a subset of the binary variables and their complements. As mentioned earlier, a clique represents a constraint in which at most one of the involved variables can be equal to one.

The main goal of the clique separation routine developed in this work is not to find \emph{a most} violated inequality, but \emph{a set of} violated inequalities. Previous work has proven this to be the best strategy. For example, \cite{Burke2012} used an algorithm to discover a most violated clique, but their computational results motivated the inclusion of additional cuts found during the separation process. This result is consistent with reports of applications of other cuts applied to different models, such as Chv\'{a}tal-Gomory cuts~\citep{Fischetti2007}. The option of inserting a large number of valid inequalities at the same time is also responsible for increasing the importance of Gomory cuts~\citep{Cornuejols2007}.

Our clique separation routine is presented in Algorithm~\ref{algClqSep}. It consists of using an LP solution $\check{x}$ and a conflict graph $G = (V, E)$ of a given MILP to separate and return a set $\mathcal{S}$ of cliques violated by solution $\check{x}$. Parameter $minViol$ is used to control the minimum violation that a clique must have to enter in $\mathcal{S}$. Our clique separator executes two main steps: it separates violated cliques in the first step (lines \ref{subgraph} to \ref{bkexe}) and extend these cliques in the second step (lines \ref{extendB} to \ref{extendE}).

\begin{algorithm}[ht]
\small
\caption{Clique Cut Separator}\label{algClqSep}
\KwIn{LP solution $\check{x}$, conflict graph $G = (V, E)$, $minViol$ and $maxCalls$.}
\KwOut{Set $\mathcal{S}$ of violated cliques.}

Let $G' = (V', E')$ be the subgraph of $G$ induced by all variables with fractional values in $\check{x}$\label{subgraph}\;

$w_{j} \gets \check{x_j}, \ \forall j \in V'$\;
$minW \gets 1 + minViol$\;
$\mathcal{S} \gets FindCliques(G', w, minW, maxCalls)$\label{bkexe}\;

\For{$C \in \mathcal{S}$} {
    \label{extendB}
    Let $d$ be the vertex in $C$ with the smallest degree\label{bkL1}\;
    $L \gets \{k \in N_{G}(d) \ | \ k \notin C\}$\label{bkL2}\;
    \While{$L \neq \emptyset$} {
        Let $l$ be the vertex in $L$ with the smallest reduced cost in the current LP relaxation\;
        Remove $l$ from $L$\;

        \If{$\forall k \in C : \ l \in N_G(k)$} {
            $C \gets C \cup \{l\}$\label{extendE}\;
        }
    }
}

\Return $\mathcal{S}$\;
\end{algorithm}

We begin generating a subgraph $G' = (V', E')$ induced by all variables (and their respective complements) with fractional values at LP solution $\check{x}$. Then, for each vertex $j$ in subgraph $G'$, we define the weight $w_j$ as the value of its corresponding variable $x_j$ in $\check{x}$. The weight of a vertex $\bar{j}$ that represents the complement of a variable $x_j$ is $w_{\bar{j}} = 1.0 - \check{x}_j$. Now we have to search for cliques in $G'$ whose sum of weights of its vertices is greater than or equal to $1 + minViol$ (line \ref{bkexe}). These are the violated cliques.

The separation of violated cliques uses a modified version of the Bron-Kerbosch (BK) algorithm~\citep{Bron1973}. Although BK has exponential computational complexity in the worst case, the use of pivoting and pruning strategies enables efficient exploration of the search space. In practice, even for harder instances, maximal cliques with high weights are found during the first stages of the search. To avoid spending too much time in the clique separation step, we limit the number of recursive calls of BK by including a parameter called $maxCalls$. Details about this algorithm are discussed in Subsection~\ref{secBK}.

After executing the BK algorithm, we have a set of violated cliques stored in $\mathcal{S}$. The clique extension module (lines \ref{extendB} to \ref{extendE}) is then performed to extend each clique $C \in \mathcal{S}$ by inserting the variables (or their complements) with integer values at the current LP solution $\check{x}$. For this, we use a greedy strategy and conflict graph $G$.

First, we create a set $L$ of candidates to enter the clique $C$. It is built with the neighbors of the vertex in $C$ with the smallest degree, excluding those that are already in $C$ (lines \ref{bkL1} and \ref{bkL2}). $N_{G}(d)$ indicates the vertices in $G$ that are adjacent to vertex $d$. Then, we try to insert additional vertices in $C$ by iteratively selecting the vertex $l \in L$ with the smallest reduced cost in the current LP relaxation. The reduced cost of a variable is the amount of penalty that would be generated if one unit of this variable was introduced into the solution. At each iteration, vertex $l$ is inserted in $C$ only if it is adjacent to all vertices in $C$. This process repeats until $L$ is empty.

\begin{figure}[ht]
    \begin{center}
        \includegraphics[width=0.4\textwidth]{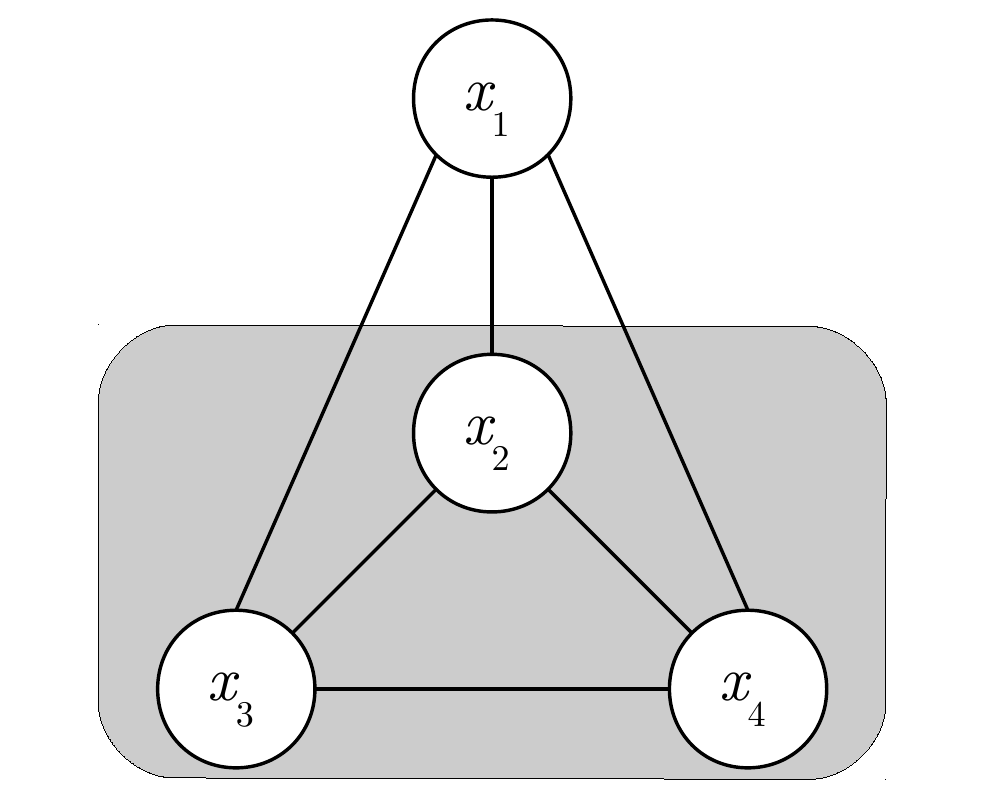}
        \caption{Example of a $K_3$ in which the extension module could be applied, transforming it into a $K_4$.}\label{figClqExt}
    \end{center}
\end{figure}

Figure~\ref{figClqExt} illustrates the importance of extending clique inequalities. Vertices within the gray area indicate variables with nonzero values in the solution of the current LP relaxation. Only vertices $x_2$, $x_3$ and $x_4$ could contribute toward defining a most violated clique inequality. Despite this, subsequent LP relaxations would include three different $K_3$ cliques, alternating the variable whose value is equal to zero. Reoptimizations of the LP could be avoided if the inequality of the $K_4$ clique was inserted immediately after the first LP relaxation of the problem. Moreover, a less dense constraint matrix may be obtained with the insertion of this dominant constraint.

\subsubsection{Bron-Kerbosch Algorithm}\label{secBK}

The main component of our clique separator is based on the \emph{Bron-Kerbosch} algorithm, which is responsible for finding cliques with weights greater than a certain threshold. BK is a backtracking-based algorithm that enumerates all maximal cliques in undirected graphs~\citep{Bron1973}.

Some strategies to improve this algorithm are present in the literature. For example, in the same work as they presented the algorithm, \cite{Bron1973} introduced a variation that employs a pivoting strategy to decrease the number of recursive calls. Following this idea, \cite{Tomita2006} proposed a pivoted version of BK where all maximal cliques are enumerated in $O(3^{\frac{|V|}{3}})$ steps. This strategy makes the pivot a vertex with the highest number of neighbors in the candidate set.

Following the idea of \cite{Tomita2006}, we implemented a pivoted version of BK. Additionally, a pruning strategy was added to accelerate the discovery of maximal cliques with weights greater than a threshold. Algorithm~\ref{algBK} details our implementation.

\begin{algorithm}[ht]
\small
\caption{Bron-Kerbosch algorithm for detecting maximal cliques with weights above a threshold.} \label{algBK}

\SetKwProg{Fn}{Function}{:}{}
\Fn{$FindCliques(G, w, minW, maxCalls)$}
{
    $R \gets \emptyset$; $P \gets V$; $X \gets \emptyset$; $\mathcal{S} \gets \emptyset$\;
    $BronKerbosch(G, w, minW, maxCalls, \mathcal{S}, R, P, X, 0)$\;
    \Return $\mathcal{S}$\;
}

\SetKwProg{Fn}{Function}{:}{}
\Fn{$BronKerbosch(G, w, minW, maxCalls, \mathcal{S}, R, P, X, numCalls)$}
{
    $numCalls \gets numCalls + 1$\;
    \If{$numCalls > maxCalls$} {
        \Return\;
    }
    \label{clqBegin}
    \If{$P \cup X = \emptyset$}
    {
    	\If{$\omega(R) \geq minW$}
    	{
    	\label{clqCheckW}
        	$\mathcal{S} \gets \mathcal{S} \cup \{R\}$\label{clqEnd}\;
    	}
    	\Return\;
    }

    \If{$\omega(R) + \omega(P) \geq minW$}
    {\label{compWEst}
        choose a pivot vertex $u \in P \cup X$\; 

        \ForEach{$v \in P \setminus N_{G}(u)$}
        { \label{recCall1}
            $BronKerbosch(G, w, minW, \mathcal{S}, R \cup \{v\}, P \cap N_{G}(v), X \cap N_{G}(v))$\label{recCall2}\;
            $P \gets P \setminus \{v\}$\label{updateP}\;
            $X \gets X \cup \{v\}$\label{updateX}\;
        }
    }
}
\end{algorithm}

Our algorithm works with three disjoint vertex sets: $R$, $P$ and $X$. Set $R$ is the set of vertices that are part of the current clique. Meanwhile, sets $P$ and $X$ are the candidate vertices to enter in $R$ and all the vertices that have already been considered in earlier steps, respectively.

The algorithm begins with $R$ and $X$ empty, while $P$ contains all the vertices of the graph. Within each recursive call, if the sets $P$ and $X$ are empty (line \ref{clqBegin}), then $R$ is a maximal clique. This clique is stored in the clique set $\mathcal{S}$ if its weight $\omega(R) = \sum_{j \in R} w_j$ is greater than or equal to the minimum weight $minW$ (lines \ref{clqCheckW} to \ref{clqEnd}).

If $R$ is not yet a maximal clique, the algorithm proceeds and calculates an upper bound to the weight that can be achieved by extending this set. This is done by adding the current weight of $R$ to the weight of candidate vertices $P$. The upper bound to the weight of $R$ is computed to avoid exploring sub-trees which would lead to cliques that do not satisfy the minimum weight $minW$ (line \ref{compWEst}).

Then, a pivot vertex $u$ is selected from $P \cup X$. It is well known that the selection of the pivot vertex is very influential on the overall performance of the method. Thus, we developed five different pivoting rules:

\begin{description}
	\item[rnd:] randomly selects a vertex.
    \item[deg:] selects the vertex with the highest degree.
    \item[wgt:] selects the vertex with the highest weight.
    \item[mdg:] selects the vertex with the highest modified degree.
    \item[mwt:] selects the vertex with the highest modified weight.
\end{description}

\noindent The modified weight of a vertex is computed as the sum of its weight and the weights of the vertices adjacent to it. To the best of our knowledge, this is the first time that different pivoting rules are evaluated for the BK algorithm in the context of clique cut separation.

Next, for each candidate vertex $v$ which is not a neighbor of pivot $u$ (line \ref{recCall1}) a recursive call is made, adding $v$ into clique $R$ and updating sets $P$ and $X$ (line \ref{recCall2}). At this point, sets $P$ and $X$ contain the neighbors of vertex $v$ which are also neighbors of the other vertices contained in clique $R$. Using this configuration, the algorithm finds all extensions of $R$ containing $v$. Once vertex $v$ has been analyzed, it is removed from $P$ and inserted into $X$ (lines \ref{updateP} and \ref{updateX}). The algorithm finishes after finding all maximal cliques with weights greater than $minW$ or when a maximum number of recursive calls $maxCalls$ is reached.

Since the most critical bottlenecks of Algorithm~\ref{algBK} are the set operations, we employ bit strings that exploit bit-level parallelism in hardware for optimizing the calculation of intersection, union and removal of sets. We encode graph $G$ as an array of bit strings, where each array entry corresponds to a row of the adjacency matrix of $G$.  The complement graph $\bar{G}$ of $G$ is also encoded as an array of bit strings to allow the implementation of efficient bitmasks operations concerning non-neighbor relations. Finally, sets $P$ and $X$ of the BK algorithm are also encoded as bit strings. With these representations, we can implement the critical set operations in Algorithm~\ref{algBK} as bitmask \emph{AND} operations~\citep{SanSegundo2018}:

\begin{itemize}
    \item $P \cap N_{G}(v)$ in line~\ref{recCall2}: \emph{AND} operation between bit string $P$ and the $v$-th row of $G$.
    \item $X \cap N_{G}(v)$ in line~\ref{recCall2}: \emph{AND} operation between bit string $X$ and the $v$-th row of $G$.
    \item $P \setminus N_{G}(u)$ in line~\ref{recCall1}: \emph{AND} operation between bit string $P$ and the $u$-th row of $\bar{G}$.
\end{itemize}

\subsection{Odd-Cycle Cut Separator}

Odd-cycle inequalities are also derived from the set packing polytope. Given a graph $G = (V, E)$, a subset $O \subseteq V$ is an odd cycle if the subgraph induced by $O$ is a simple cycle with an odd number of vertices. In this case, the subgraph must have $|O|$ adjacent edges such that each vertex is incident to exactly two vertices. Thus, an odd cycle $O$ formed by a set of binary variables (or their complements) defines the odd-cycle inequality:

\begin{equation*}
    \sum_{j\in O}x_{j} \leq \frac{|O| - 1}{2}
\end{equation*}

\noindent This inequality ensures that at most half of the variables can be activated.

Our odd-cycle separation routine is described in Algorithm~\ref{algOddSep}. It is based on the concepts presented by \cite{Rebennack2009} and returns a set $\mathcal{W}$ of tuples containing the violated odd cycles and their respective wheel centers. First, an auxiliary bipartite graph~$G' = (V', E')$ is created from the original conflict graph $G = (V, E)$. Lines \ref{createSubgB} to \ref{createSubgE} present the creation of $G'$. The vertex set $V'$ is formed by two subsets $V_1$ and $V_2$: for each vertex $j \in V$, two vertices $j_1$ and $j_2$ are created in $V'$, where $j_1 \in V_1$ and $j_2 \in V_2$. Additionally, for each edge $(j, k) \in E$, two edges $(j_1, k_2)$ and $(j_2, k_1)$ are inserted into $E'$, where $j_1, k_1 \in V_1$ and $j_2, k_2 \in V_2$. The auxiliary graph $G'$ is bipartite since there is no edges connecting two vertices of $V_1$ or two vertices of $V_2$.

\begin{algorithm}[htbp]
\small
\caption{Odd-Cycle Cut Separator}\label{algOddSep}
\KwIn{LP solution $\check{x}$ and conflict graph $G = (V, E)$.}
\KwOut{Set $\mathcal{W}$ of tuples containing violated odd cycles and their respective wheel centers.}
$\mathcal{W} \gets \emptyset$\;
$V' \gets \{j_1, j_2 \ | \ j \in V\}$\label{createSubgB}\;
$E' \gets \{(j_1, k_2), (j_2, k_1) \ | \ (j, k) \in E\}$\;
$G' \gets (V', E')$\label{createSubgE}\;
\For{$(j, k) \in E$} {
    \label{defWOddB}
    $w(j_1, k_2) = (1 - \check{x}_j - \check{x}_k) / 2$\;
    $w(j_2, k_1) = (1 - \check{x}_j - \check{x}_k) / 2$ \label{defWOddE}\;
}

\For{$j \in V$} {
    $P \gets ShortestPath(j_1, j_2, G', w)$\;
    Convert path $P$ to an odd cycle $O$ in the original graph $G$\label{createO}\;
    Let $\tilde{E}$ be the edge set of the subgraph of $G$ induced by $O$\;
    $cost \gets 0$\;
    \For{$(j, k) \in \tilde{E}$} {
        $cost \gets cost + w(j, k)$;\
    }

    \If{$cost < 0.5$} {
    	$C \gets \emptyset$\label{oddExtB}\;
        Let $d$ be the vertex in $O$ with the smallest degree\label{oddL1}\;
		$L \gets \{k \in N_{G}(d) \ | \ k \notin O, \ \exists k \in N_{G}(j) \ \forall j \in O\}$\label{oddL2}\;
		\While{$L \neq \emptyset$} {
		    Let $l$ be the vertex that corresponds to the variable with the smallest reduced cost in $L$\;
		    Remove $l$ from $L$\;

		    \If{$\forall k \in C : \ l \in N_G(k)$} {
		        $C \gets C \cup \{l\}$\label{oddExtE}\;
		    }
		}
		$\mathcal{W} \gets \mathcal{W} \cup \{(O, C)\}$\label{oddInsert}\;
    }
}
\Return $\mathcal{W}$\;
\end{algorithm}

The next step is to compute the weight of each edge of $G'$ (lines \ref{defWOddB} to \ref{defWOddE}), since our cut separator works with an edge-weighted graph. The weight of each edge in the auxiliary graph $G'$ is computed according to the corresponding edge $(j, k)$ of the original graph $G$, which is defined as:

\begin{equation*}
    w(j, k) = \frac{1 - \check{x}_j - \check{x}_k}{2}
\end{equation*}

\noindent where $\check{x}_j$ and $\check{x}_k$ are the values of variables $x_j$ and $x_k$ at LP solution $\check{x}$. Here, variables $x_j$ and $x_k$ are conflicting, which implies that $\check{x}_j + \check{x}_k \leq 1$ and, consequently, $w(j, k) \geq 0$.

After creating the auxiliary bipartite graph and computing its edge weights, the search for violated odd cycles begins. For each vertex $j \in V$ we run Dijkstra's algorithm in $G'$ to find the shortest path $P$ from $j_1$ to $j_2$. The shortest path has an odd number of edges since vertices $j_1$ and $j_2$ are in two different sets of the bipartition. Then, the corresponding odd-cycle $O$ is constructed from the shortest path $P$ (line \ref{createO}). The resulting odd-cycle inequality is violated by the current LP solution $\check{x}$ if and only if:

\begin{equation*}
    \sum_{(j, k) \in \tilde{E}} w(j, k) < 0.5
\end{equation*}

\noindent where $\tilde{E}$ is the set of edges of the subgraph of $G$ induced by the variables in odd cycle $O$. Thus, Algorithm~\ref{algOddSep} tries to find one odd-cycle inequality (namely, a most violated one) for each variable. Odd-cycles of size three are discarded since they correspond to cliques and could be found by the clique cut separation procedure.

When a violated odd cycle is found, a lifting step is performed (lines \ref{oddExtB} to \ref{oddExtE}). This step tries to transform the violated odd cycle into an odd wheel. In graph theory, an odd wheel is an odd cycle that contains an additional vertex that is adjacent to all other vertices. Thus, an odd wheel can be obtained by inserting a variable into the center of the odd cycle. An odd-wheel inequality has the following format:

\begin{equation*}
    \sum_{j\in O} x_{j} + \frac{|O| - 1}{2} x_c \leq \frac{|O| - 1}{2}
\end{equation*}

\noindent where $O$ is an odd cycle and $x_c$ is a variable that has conflict with all of those in $O$.

Our lifting step consists of a new approach: we use a greedy strategy that finds and inserts a clique $C$ in the center of the odd cycle. In the literature, it is common to consider the insertion of only one variable into the center of the odd cycle, such as the lifting strategy presented by \cite{Rebennack2009}. Initially, we select a vertex $d$ with the smallest degree between the vertices of $O$. Then, we create a set $L$ of candidates to compose clique $C$, containing the neighbors of $d$ that are conflicting with all vertices in $O$ but are not included in $O$ (lines \ref{oddL1} and \ref{oddL2}). Following, we construct clique $C$ by iteratively selecting the vertex $l \in L$ whose corresponding variable has the smallest reduced cost. A vertex $l$ is inserted in $C$ only if it is adjacent to all vertices in $C$. This process repeats until $L$ is empty. Finally, a tuple formed by violated odd-cycle $O$ and its corresponding wheel center $C$ is inserted in $\mathcal{W}$ (line \ref{oddInsert}).

\begin{figure}[ht]
    \begin{center}
        \includegraphics[width=0.4\textwidth]{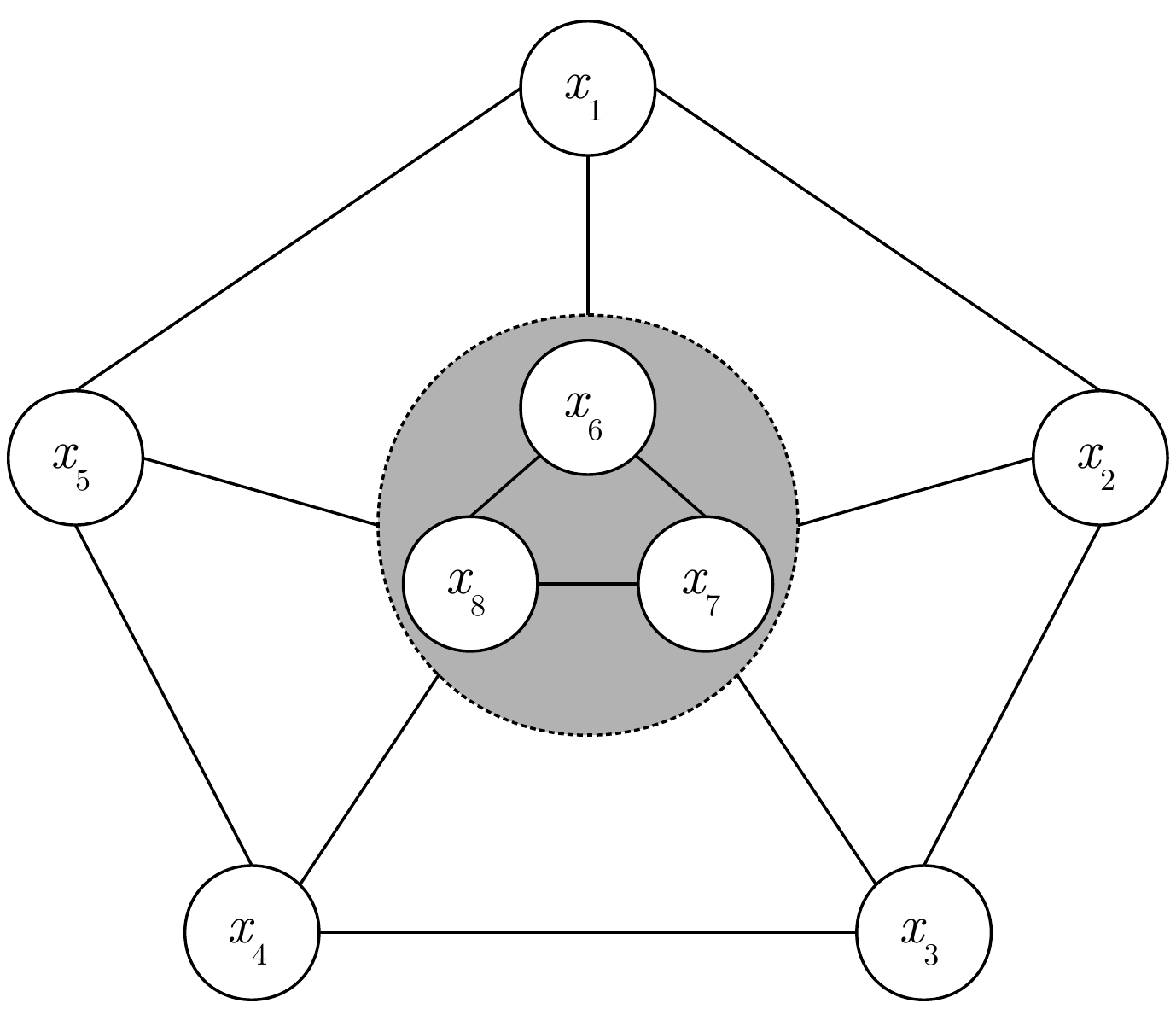}
        \caption{Example of an odd cycle with the inclusion of a wheel center. Vertices $x_6$, $x_7$ and $x_8$ are connected to all the vertices of the odd cycle formed by $\{x_1, x_2,x_3, x_4, x_5\}$.}\label{figOdd}
    \end{center}
\end{figure}

Figure~\ref{figOdd} illustrates an odd wheel formed by the inclusion of the clique involving variables $\{x_6, x_7, x_8\}$ into the center of the odd cycle formed by variables $\{x_1, x_2,x_3, x_4, x_5\}$. In this example, each variable in the clique is conflicting with all variables that compose the odd cycle. The odd-wheel inequality associated with the odd cycle of this figure is: 

\begin{equation*}
    x_1 + x_2 + x_3 + x_4 + x_5 + 2x_6 + 2x_7 + 2x_8 \leq 2
\end{equation*}

\section{Computational Results}\label{experiments}

This section presents the results of computational experiments conducted to evaluate the performance of the proposed conflict graph-based algorithms and data structures. All computational experiments were carried out on four computers with Intel Core i7-4790 3.60 GHz processors and 32 GB of RAM running Ubuntu Linux version 18.04 64-bit. The source code was developed in C++ programming language and compiled with g++ version 7.4.0 using the optimization level \texttt{Ofast}. The algorithms and data structures were included in the development version of the CBC solver, which will become CBC 3.

\subsection{Instance Sets}

The instances used in the experiments consist of $320$ MILPs divided into five groups:

\begin{description}
	\item[bmc:] instances of Bandwidth Multicoloring Problem~\citep{Dias2016};
	\item[bpwc:] instances of Bin Packing Problem with Conflicts~\citep{Sadykov2013};
	\item[miplib:] instances of the current and previous versions of the Mixed-Integer Problem Library (MIPLIB) benchmark set~\citep{Miplib2017};
	\item[rostering:] instances of Nurse Rostering problem~\citep{Haspeslagh2014};
	\item[timetabling:] instances of Educational Timetabling~\citep{Fonseca2017}.
\end{description}

Table~\ref{tabInstSummary} contains summarized information concerning the instance sets. In this table, column ``size'' presents the number of instances of each instance set and ``cols'' contains the average number of variables. Columns ``int'', ``bin'' and ``con'' present, respectively, the average number of integer, binary and continuous variables of each instance set. Finally, columns ``rows'' and ``nz'' detail information with respect to the average number of constraints and nonzeros coefficients of each instance set.

\begin{table}[ht]
\scriptsize
\caption{Characteristics of the instance sets used in the experiments.}\label{tabInstSummary}
\begin{center}
\begin{tabular}{lrrrrrrr}
\toprule
\mc{group} & \mc{size} & \mc{cols} & \mc{int} & \mc{bin} & \mc{con} & \mc{rows} & \mc{nz} \\
\midrule
bmc & 9 & 15,606.33 & 0.00 & 15,605.33 & 1.00 & 398,899.89 & 813,363.11 \\
bpwc & 20 & 13,223.40 & 0.00 & 13,223.40 & 0.00 & 148,569.05 & 322,656.10 \\
miplib & 253 & 38,603.17 & 343.98 & 24,953.26 & 13,305.94 & 44,159.98 & 515,170.28 \\
rostering & 22 & 35,054.77 & 4.45 & 35,050.32 & 0.00 & 14,082.41 & 626,280.73 \\
timetabling & 16 & 20,768.25 & 10,828.75 & 9,939.50 & 0.00 & 40,902.94 & 159,929.25 \\
\bottomrule
\end{tabular}
\end{center}
\end{table}

\subsection{Building Conflict Graphs}

The first experiment was conducted to compare the performance of our algorithm for building CGs, named as \emph{ICE}, against the pairwise inspection scheme of \cite{Atamturk2000}, referred to here as \emph{PI}, and the clique extraction algorithm of \cite{Achterberg2007}, denoted as \emph{CE}. For this purpose, we considered the execution times and memory usage in creating the graphs. \emph{PI} and \emph{CE} were implemented based on the descriptions given in their respective works. Since \emph{PI} only detects pairs of conflicts, we use an adjacency list for each vertex of the graph. The conflict storage of \emph{CE} uses the same data structure employed in our algorithm.

In this experiment, \emph{PI} failed to construct graphs for eight instances: \emph{eilA101-2}, \emph{eilB101.2}, \emph{eilD76.2}, \emph{nw04}, \emph{s100}, \emph{square41}, \emph{square47} and \emph{supportcase6}. \emph{PI} needed more than $32$ GB of memory to construct and store CGs for these instances. They have some set packing and set partitioning constraints formed by a large number of variables, whose pairwise storage of conflicts results in excessive memory consumption. Since the other algorithms can explicitly store cliques, they did not face memory issues with respect to these instances. We penalize the cases where \emph{PI} cannot construct CGs due to memory limitation, assigning for each affected instance a memory usage of $32$ GB and an execution time of $1{,}800$ seconds.

\begin{figure}[htbp]
\centering
\begin{subfigure}{.45\textwidth}
  \centering
  \includegraphics[width=1.0\linewidth]{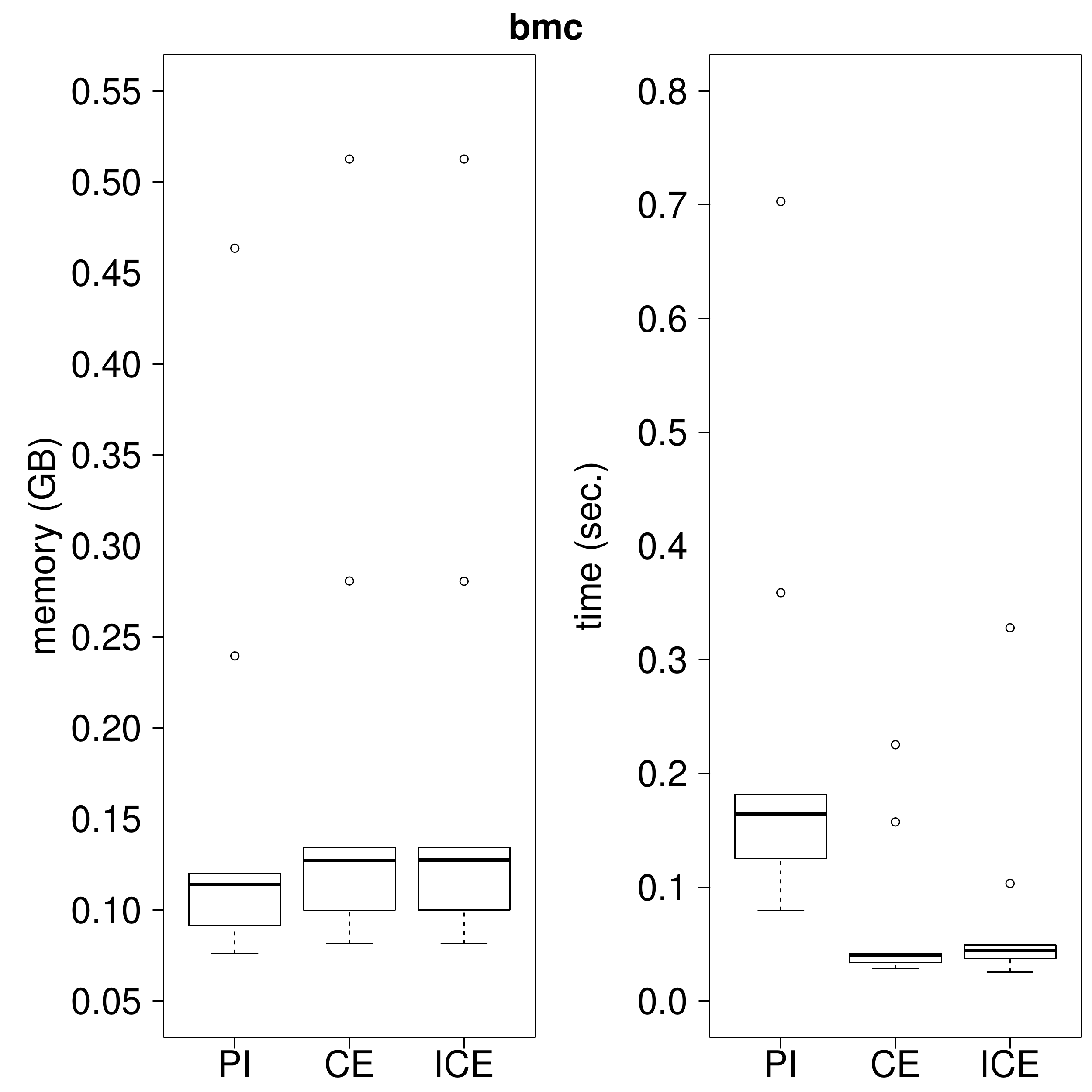}
\end{subfigure}
\begin{subfigure}{.45\textwidth}
  \centering
  \includegraphics[width=1.0\linewidth]{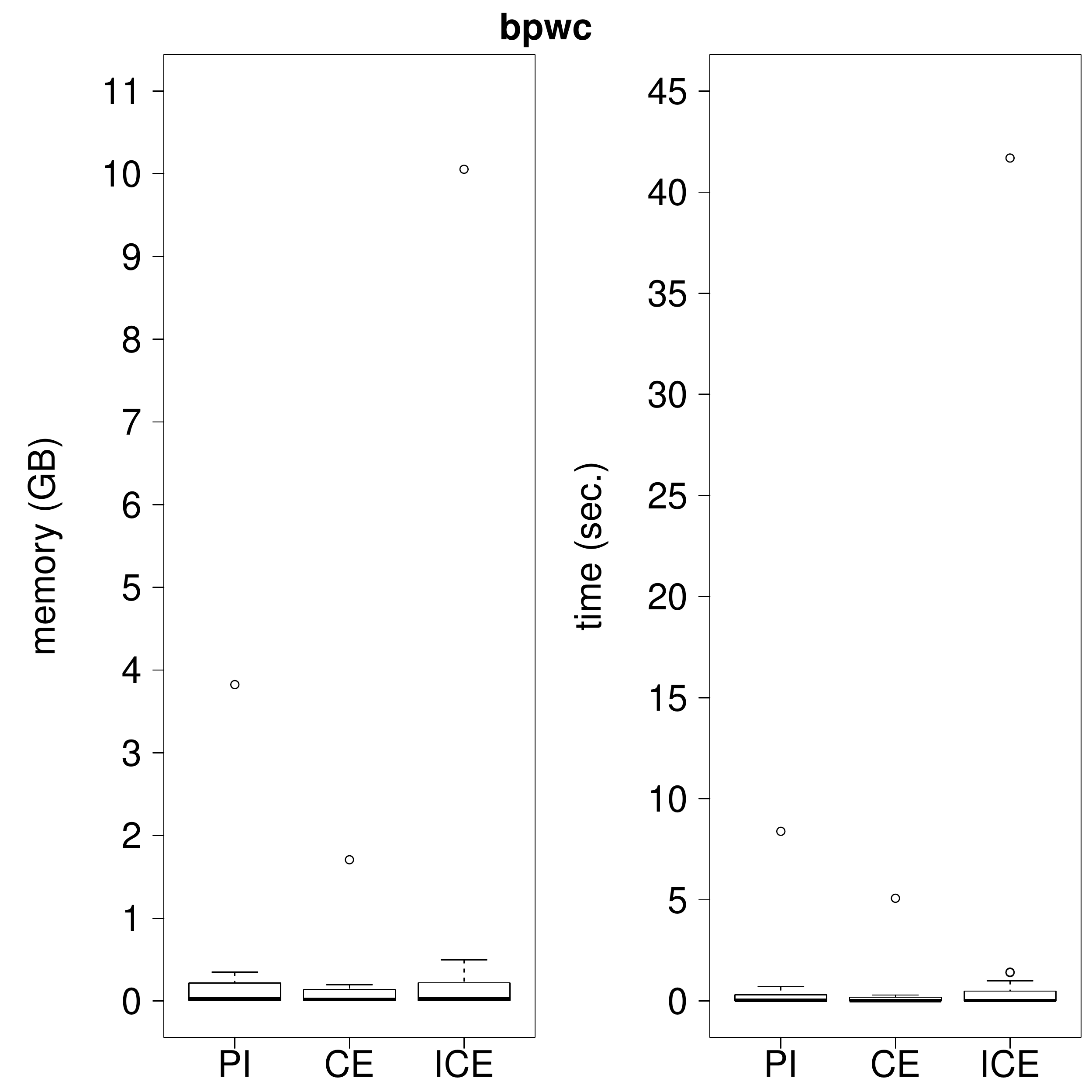}
\end{subfigure}
\par\bigskip
\begin{subfigure}{.45\textwidth}
  \centering
  \includegraphics[width=1.0\linewidth]{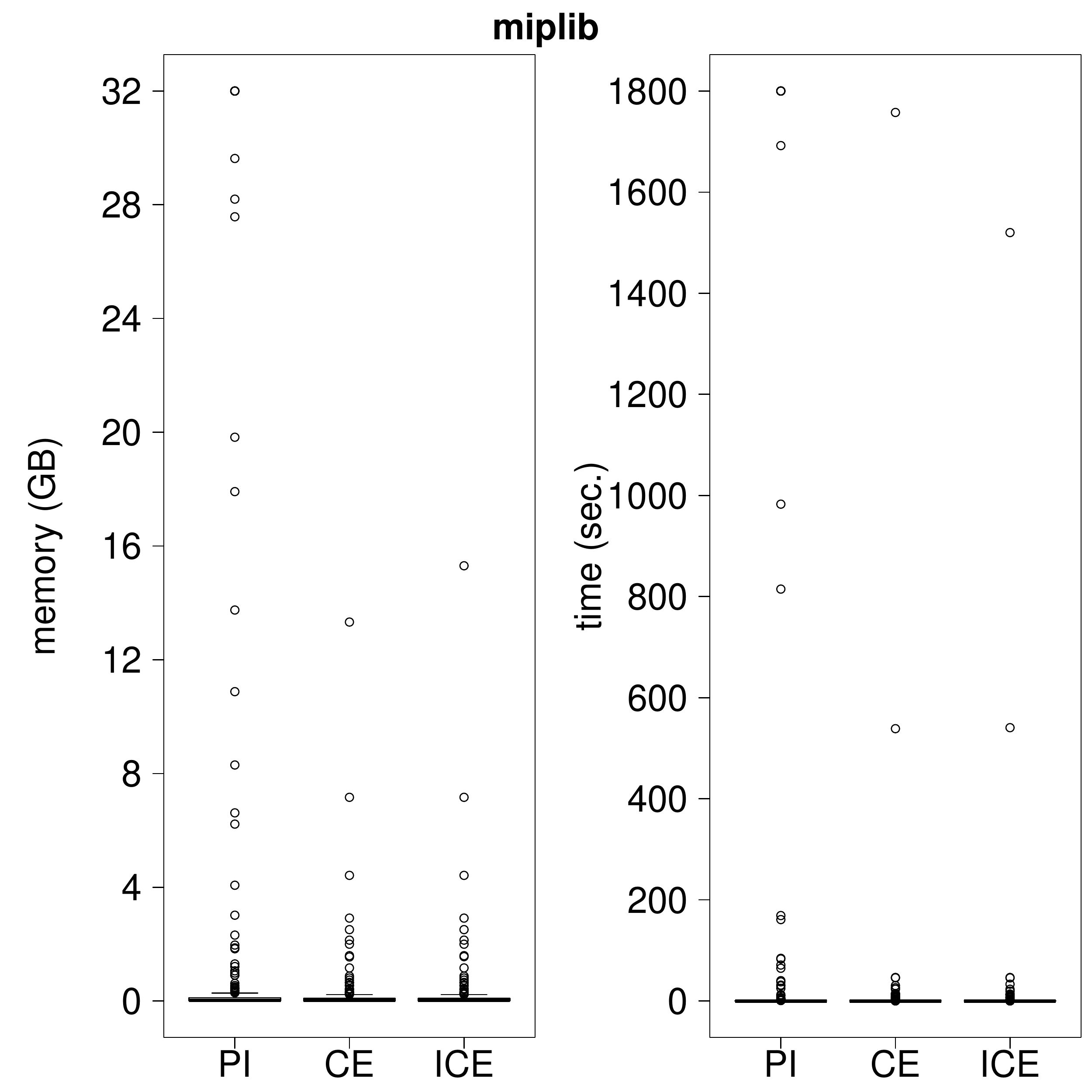}
\end{subfigure}
\begin{subfigure}{.45\textwidth}
  \centering
  \includegraphics[width=1.0\linewidth]{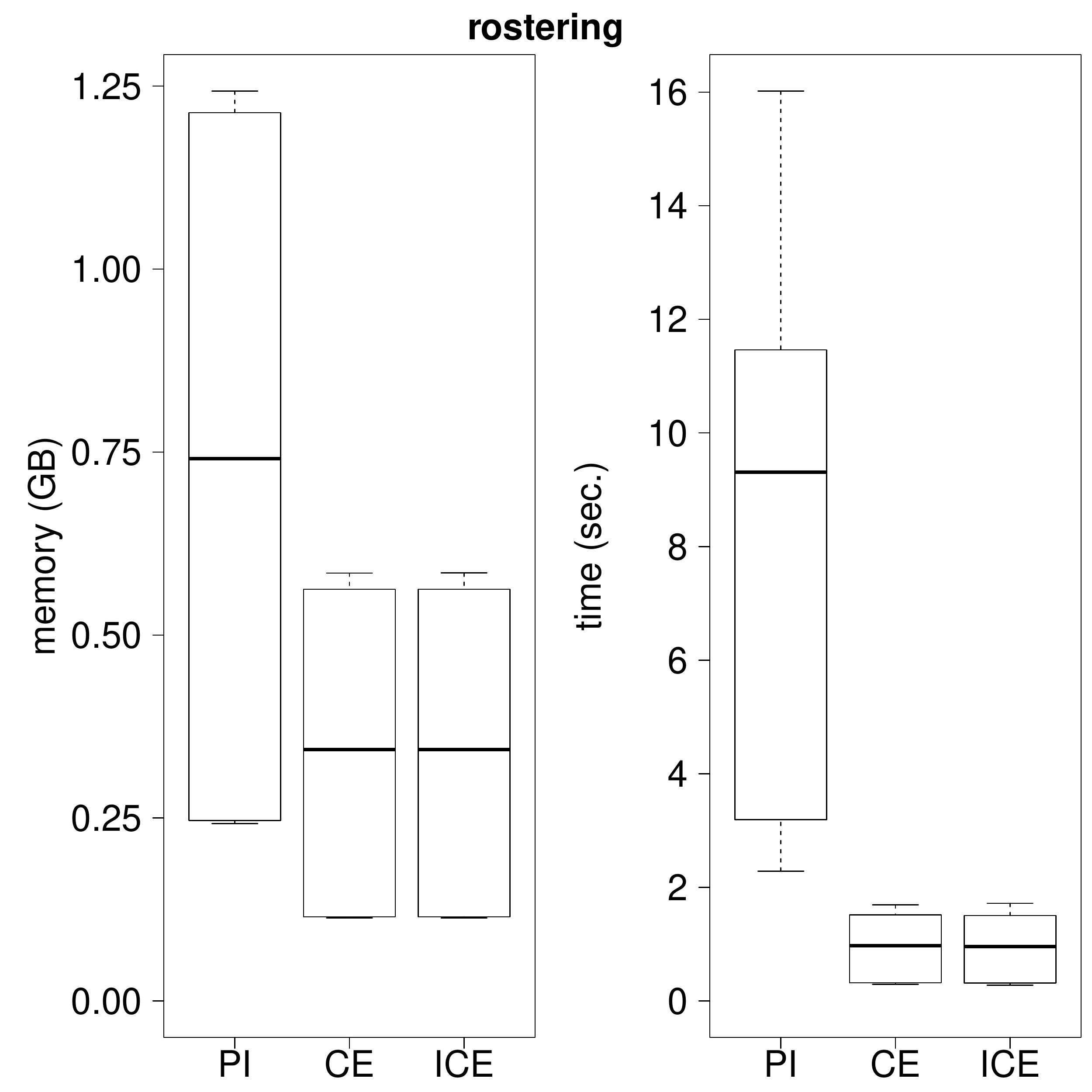}
\end{subfigure}
\par\bigskip
\begin{subfigure}{.45\textwidth}
  \centering
  \includegraphics[width=1.0\linewidth]{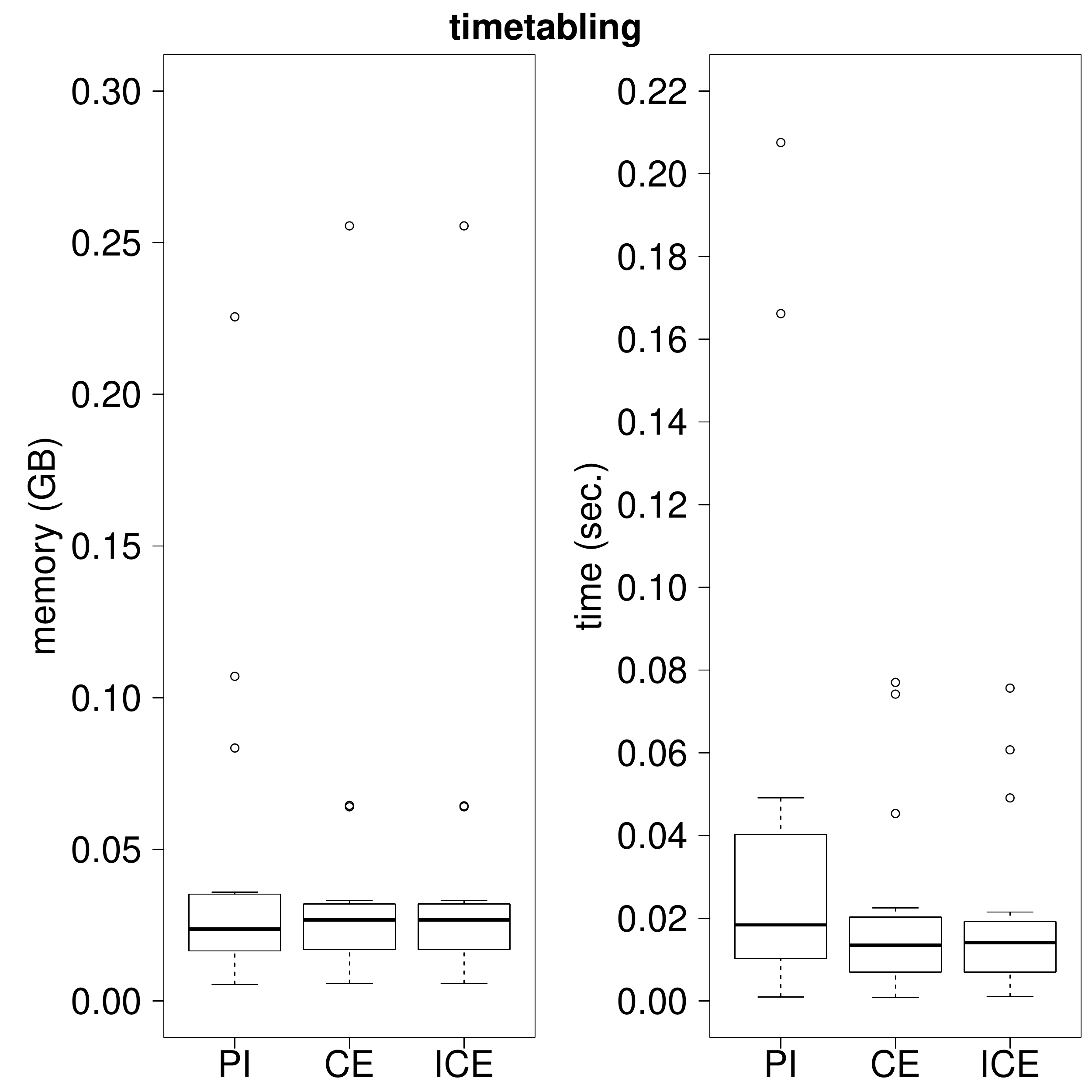}
\end{subfigure}
\caption{Execution times and memory usage in the construction of CGs.}\label{figBuildCG}
\end{figure}

Figure~\ref{figBuildCG} shows the memory usage and time spent in constructing CGs for each algorithm and each instance set. The algorithms presented similar memory usage in instances of \emph{bmc} and \emph{timetabling}. Although the execution times are less than $1$ second for these instances, \emph{CE} and \emph{ICE} obtained values that are smaller than \emph{PI}.

Memory usage and execution times are similar in instance set \emph{bpwc}, except for instance \emph{uELGN\_BPWC\_3\_2\_18}. In this instance, \emph{ICE} found about $8.7$ million additional conflicts. The higher number of conflicts detected implied a greater consumption of time and memory.

The greatest performance gain with the use of the clique extraction approach and our data structures was obtained on the \emph{miplib} dataset. For some instances of this set, several cliques were detected and explicitly stored, contributing to the significative reduction of execution time and memory consumption on the construction of CGs. For example, \emph{PI} needs more than $32$ GB of memory to construct a graph for instance \emph{eilD76.2}, while \emph{CE} and \emph{ICE} were able to build graphs for the same instance using only $52.10$ MB. The results obtained in instance set \emph{rostering} also demonstrates that \emph{CE} and \emph{ICE} are more efficient than \emph{PI} in terms of memory consumption and time spent to create CGs.

In general, the combination of the strategy to avoid analyzing constraints that would not lead to the discovery of conflicts with the efficient clique extraction approach and the use of our optimized data structures contributed to decrease the amount of time and memory required to construct CGs. \emph{ICE} built and stored CGs for the considered instance problems spending, on average, $252.47$ MB of memory and $7.60$ seconds. In comparison with \emph{PI}, these results represent a decrease of $85.66\%$ in memory consumption and $87.20\%$ in the execution time.

The execution time and memory consumption of \emph{CE} and \emph{ICE} were similar, except in instances where \emph{ICE} found more conflicts. Table~\ref{tabCGNewConfs} presents $27$ instances in which \emph{ICE} found more conflicts than \emph{CE}, quantifying the improvement obtained. The number of new conflicts refers to the number of new edges that were included in the conflict graphs.

\begin{table}[htbp]
\scriptsize
\caption{Instances where new conflicts were discovered.}\label{tabCGNewConfs}
\begin{center}
\begin{tabular}{lcrrrr}
\toprule
\mc{instance} & \mc{group} & \mc{CE} & \mc{ICE} & \mc{new conflicts} & \mc{\% increase}\\ 
\midrule
n2seq36q & miplib & 20,653,320 & 20,653,344 & 24 & $<$ 0.01 \\
p0548 & miplib & 920 & 980 & 60 & 6.52 \\
istanbul-no-cutoff & miplib & 72 & 164 & 92 & 127.78 \\
p2756 & miplib & 5,640 & 5,732 & 92 & 1.63 \\
ua\_BPWC\_1\_9\_2 & bpwc & 13,506 & 13,718 & 212 & 1.57 \\
ta\_BPWC\_6\_9\_8 & bpwc & 13,420 & 13,868 & 448 & 3.34 \\
ta\_BPWC\_5\_7\_4 & bpwc & 12,876 & 13,474 & 598 & 4.64 \\
ta\_BPWC\_5\_7\_1 & bpwc & 12,736 & 13,440 & 704 & 5.53 \\
ua\_BPWC\_1\_8\_10 & bpwc & 48,532 & 49,798 & 1,266 & 2.61 \\
ta\_BPWC\_5\_5\_5 & bpwc & 31,198 & 33,826 & 2,628 & 8.42 \\
tELGN\_BPWC\_6\_8\_9 & bpwc & 28,542 & 31,598 & 3,056 & 10.71 \\
uMIMT\_BPPC\_2\_9\_1 & bpwc & 58,716 & 64,978 & 6,262 & 10.66 \\
neos-631694 & miplib & 377,746 & 385,368 & 7,622 & 2.02 \\
tELGN\_BPWC\_6\_6\_20 & bpwc & 78,628 & 87,642 & 9,014 & 11.46 \\
uELGN\_BPWC\_3\_9\_18 & bpwc & 253,020 & 281,838 & 28,818 & 11.39 \\
neos-631784 & miplib & 7,918,996 & 7,966,118 & 47,122 & 0.60 \\
neos-662469 & miplib & 2,401,182 & 2,460,118 & 58,936 & 2.45 \\
tMIMT\_BPPC\_6\_3\_4 & bpwc & 542,740 & 629,266 & 86,526 & 15.94 \\
supportcase18 & miplib & 1,913,864 & 2,023,850 & 109,986 & 5.75 \\
tELGN\_BPWC\_7\_6\_16 & bpwc & 1,106,790 & 1,241,972 & 135,182 & 12.21 \\
neos-631709 & miplib & 18,519,540 & 18,777,268 & 257,728 & 1.39 \\
uMIMT\_BPPC\_2\_5\_2 & bpwc & 2,636,516 & 2,978,096 & 341,580 & 12.96 \\
tMIMT\_BPPC\_8\_7\_5 & bpwc & 3,960,040 & 4,469,682 & 509,642 & 12.87 \\
uMIMT\_BPPC\_3\_7\_6 & bpwc & 4,177,466 & 4,713,606 & 536,140 & 12.83 \\
ta\_BPWC\_7\_1\_8 & bpwc & 5,488,238 & 6,404,428 & 916,190 & 16.69 \\
neos-631710 & miplib & 129,069,680 & 130,707,306 & 1,637,626 & 1.27 \\
uELGN\_BPWC\_3\_2\_18 & bpwc & 53,896,940 & 62,638,998 & 8,742,058 & 16.22 \\
\bottomrule
\end{tabular}
\end{center}
\end{table}

Our algorithm detected more conflicts than \emph{CE} in constraints whose minimum and maximum coefficients of the variables are different and the highest coefficients are close to the RHS of the constraints. Instances of Bin Packing Problem with Conflicts use several constraints with this characteristic to model the capacity of the bins. It is the case of instance \emph{uELGN\_BPWC\_3\_2\_18}, where more than $8.7$ million of new edges were inserted in the associated CG. The highest increase of conflicts, in percentage, was obtained in instance \emph{istanbul-no-cutoff}, in which \emph{ICE} is responsible for detecting $127.78\%$ more conflicts.

\subsection{Clique Strengthening}

The next experiment evaluated the ability of our preprocessing routine to produce strengthened formulations and to reduce the size of MILPs with respect to the number of constraints. To assess this, we ran our clique strengthening routine for all instances of the instance sets, limiting the execution of the algorithm to constraints with at most $128$ variables (i.e., $\alpha_{max} = 128$). This value was defined on a preliminary experiment that investigated the impact of setting different values to $\alpha_{max}$ in clique strengthening routine, considering the execution time and the improvement in the LP relaxation of the MILPs.

After performing clique strengthening, we used the \emph{COIN-OR Linear Program Solver} (CLP)\footnote{\url{https://github.com/coin-or/Clp}} to solve the LP relaxation, and then we calculated the gap closed. The percentage of the integrality gap closed is computed as follows:

\begin{equation}\label{gapClosed}
	gapClosed = 100 - 100 \times \frac{bestSol - currentLP}{bestSol - firstLP},
\end{equation}

\noindent where $bestSol$ is the best-known solution of the MILP, $firstLP$ is the objective value of the solution of the root node LP relaxation and $currentLP$ represents the value of the final LP relaxation.

\begin{figure}[ht]
\centering
\begin{subfigure}{.45\textwidth}
  \centering
  \includegraphics[width=1.0\linewidth]{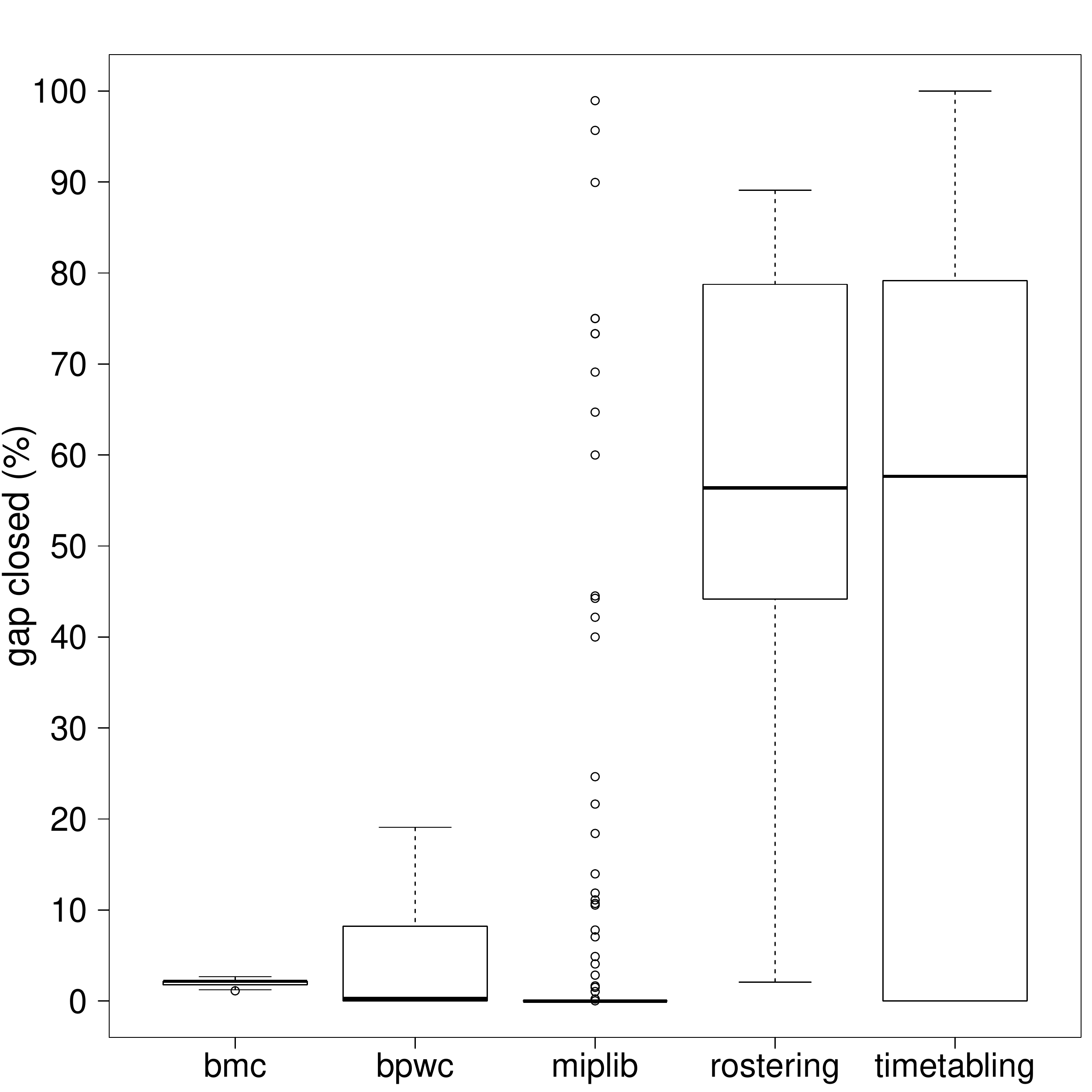}
\end{subfigure}
\begin{subfigure}{.45\textwidth}
  \centering
  \includegraphics[width=1.0\linewidth]{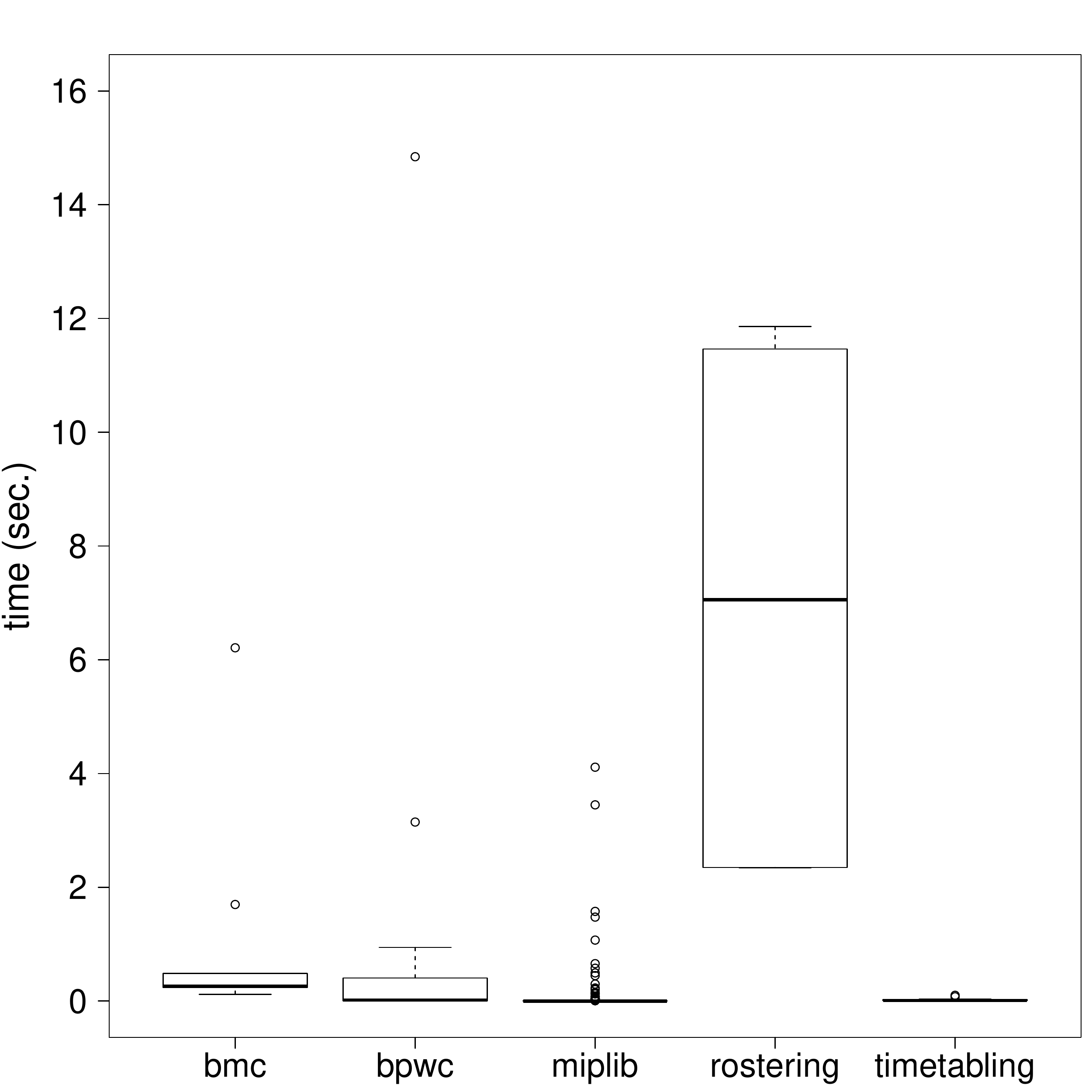}
\end{subfigure}
\begin{subfigure}{.45\textwidth}
  \centering
  \includegraphics[width=1.0\linewidth]{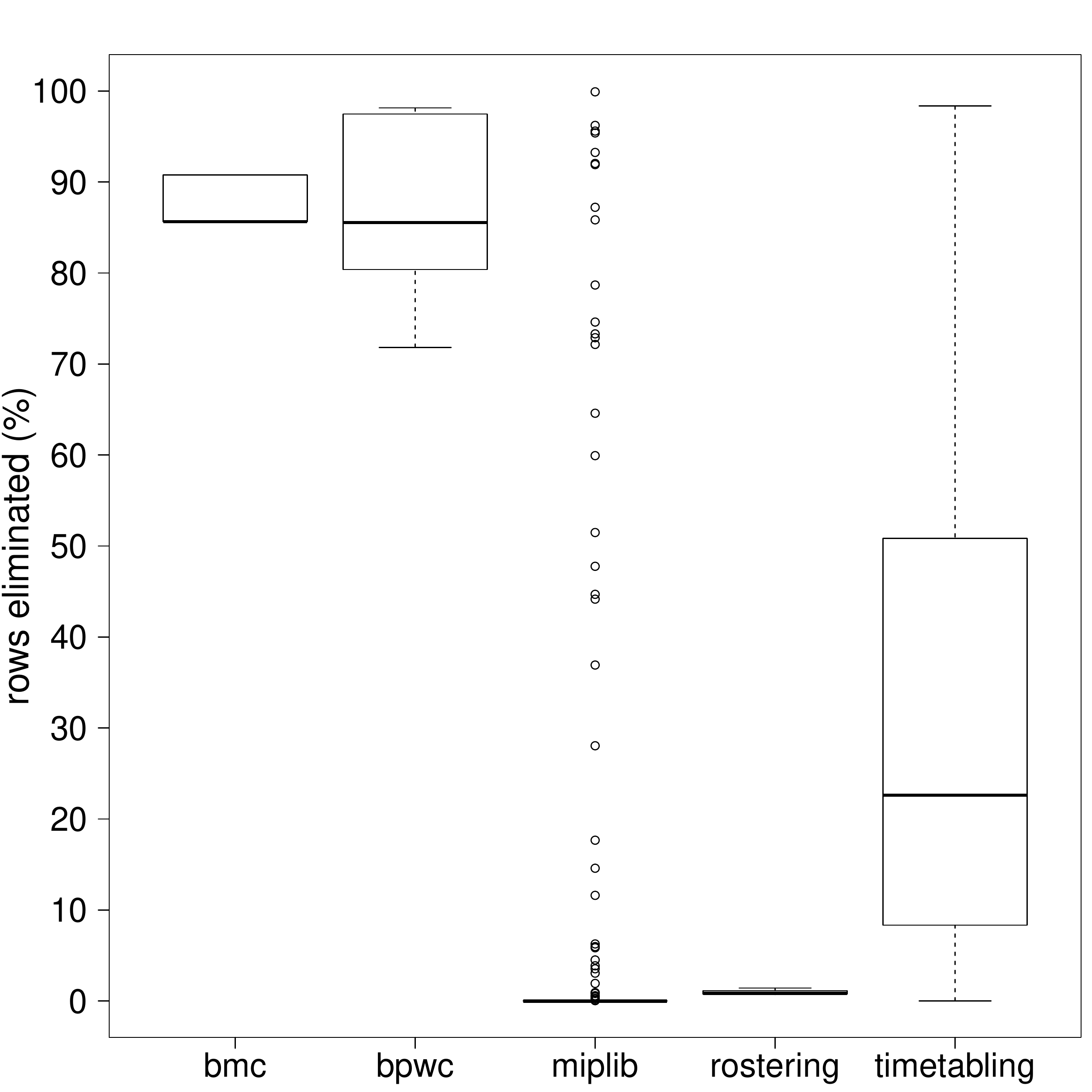}
\end{subfigure}
\caption{Results of the execution of the clique strengthening routine.}\label{figClqStr}
\end{figure}

Figure~\ref{figClqStr} provides the results regarding the percentage of rows eliminated, execution time, and gap closed by executing the clique strengthening routine in the considered instance sets. Despite the low impact on improving the values of the LP relaxation, there was a significant reduction in the number of constraints of the MILPs from \emph{bmc} and \emph{bpwc}. The percentage of constraints eliminated in instances from set \emph{bmc} was more than 85\% and above 71\% in instances from \emph{bpwc}.

The execution of the preprocessing routine neither improved the linear relaxation nor reduced the number of constraints for $205$ of $253$ instances from \emph{miplib}. Clique strengthening did not affect these instances since they have no set packing constraints. For the other $48$ instances of this group, our preprocessing routine was able to close the gap by up to 99\% and reduce the number of constraints by up to 99.91\%. Instance \emph{sorrel3}, for example, had its number of constraints reduced by 95.39\% and its integrality gap was closed by $98.94\%$. This instance belongs to the Maximum Independent Set Problem and contains $169{,}162$ set packing constraints of size two that are used to model the edges of a graph.

Several constraints of the instances from set \emph{rostering} were extended, but few became dominated. Thus, there was no reduction in the number of constraints. However, the execution of the clique strengthening routine in these instances allowed a significant improvement in the values of LP relaxations, closing the gap by up to 89\%.

A significant improvement was also observed in the value of the LP relaxation on instances of set \emph{timetabling}. The preprocessing routine was able to reduce the number of constraints of instances \emph{trd445c} and \emph{trdta0010} in $98\%$ and $80\%$, respectively. Moreover, the integrality gaps in these instances were completely closed. These instances belong to a real Educational Timetabling problem of a Brazilian university~\citep{Goncalves2008}, containing a large number of set packing constraints that are used to model pairs of conflicting assignments based on student enrollments.

In general, the time spent to run the clique strengthening routine was small, even for the cases where several constraints were extended and removed. Considering all instance problems, the maximum execution time of this routine was $14.84$ seconds. Furthermore, the execution time of the preprocessing routine was less than one second for $289$ of $320$ instances. Most instances whose execution times are greater than one second belong to set \emph{rostering}. Instances of this set have thousands of set packing constraints that are not dominated by the extended constraints. Consequently, the clique extension procedure was performed in almost all of the set packing constraints of these instances.

As the results show, clique strengthening can both reduce the number of constraints and also produce stronger formulations. It is more effective when applied to MILPs that have several constraints expressed by pairs of conflicting variables.

\subsection{Pivoting Rules of Bron-Kerbosch Algorithm}

We conducted computational experiments to evaluate the performance of the pivoting rules that we implemented in the BK algorithm (Section~\ref{secBK}). First, we generated an instance set containing several vertex-weighted graphs. We ran our clique cut separator in each instance problem, stopping this routine when no cut was separated or after performing three iterations of the cut generation loop. The pivoting rule of BK was randomly selected at each execution of the algorithm. In each iteration of the cut generation loop, we used CLP to solve the LP relaxation of the MILPs and saved the subgraphs induced by the variables with fractional values. After performing these steps, $438$ conflict graphs were generated.

Then, we ran the BK algorithm to detect the violated cliques in these graphs, limiting the maximum number of recursive calls to $100{,}000$ ($maxCalls = 100{,}000$). We investigated the performance of five versions of the BK algorithm that differ only from the pivoting rule. These versions are referred to here according to their pivoting rules, which were presented in Section~\ref{secBK}.

Table~\ref{tabBKSummary} presents the summarized results for the execution of each version of the BK algorithm. In this table, column ``exact'' indicates the number of graphs for which the algorithm ran completely, without stopping for the maximum number of recursive calls. Column ``avg calls'' presents the average number of recursive calls made by the algorithm. The average and the maximum number of violated cliques found are presented in columns ``avg clqs'' and ``max clqs''. Finally, columns ``avg time'' and ``max time'' present the average and the maximum time spent, in seconds, by each version of the BK algorithm.

\begin{table}[ht]
\small
\caption{Summarized results for the execution of BK algorithm with different pivoting rules.}\label{tabBKSummary}
\begin{center}
\begin{tabular}{lrrrrrr}
\hline
\mc{version} & \mc{exact} & \mc{avg calls} & \mc{avg clqs} & \mc{max clqs} & \mc{avg time} & \mc{max time} \\ \hline
rnd & 416 & 12,515.72 & 370.90 & 8,940 & 0.57 & 35.24\\
deg & 414 & 13,506.78 & 356.14 & 11,559 & 0.56 & 26.95\\
wgt & 424 & 9,759.37 & 381.66 & 11,115 & 0.28 & 14.80\\
mdg & 413 & 13,393.91 & 355.09 & 11,660 & 0.49 & 26.64\\
mwt & 410 & 13,790.83 & 357.18 & 9,076 & 0.53 & 34.09\\
\bottomrule
\end{tabular}
\end{center}
\end{table}

The execution of the BK algorithm was very fast. Regardless of the pivoting rule, the time spent by this algorithm was less than one second for $94\%$ of the instances. The instances in which the algorithm spent more than one second have dense conflict graphs with many cliques explicitly stored. Consequently, the process of iterating over the conflicts to encode the graphs as arrays of bit strings took the largest portion of the execution times in these instances. 

According to the results, the pivoting rule that defines the vertex with the highest weight as the pivot obtained the best results. The number of recursive calls made by this version is up to $29\%$ less than those made by other versions. The reduction in the number of recursive calls implied a decrease in the execution time, making \emph{wgt} the fastest version among those tested. In addition, \emph{wgt} ran completely for a greater number of instances and found more cliques than the other versions. Based on these results, we defined the pivoting rule of \emph{wgt} as the default pivoting rule of our implementation of the BK algorithm.

\subsection{Clique Cut Separator}

After choosing the pivoting rule to be used in the BK algorithm, we evaluated the ability of our clique cut separator in tightening the LP relaxations. In this experiment, we considered two versions of our clique cut separator: one version with the lifting module disabled and other with this module activated. These versions are referred to here as \emph{bkclq} and \emph{bkclqext}, respectively.

We compared the performance of our clique cut separator against the clique cut separators implemented in three MILP solvers. The first clique cut separator that we compared, referred to here as \emph{cglclq}, is used by the COIN-OR CBC solver and provided by the COIN-OR Cut Generation Library (CGL)\footnote{\url{https://github.com/coin-or/Cgl}}. We used the C++ API of CGL to develop a routine that calls the clique cut separator at each iteration of the cut generation loop.

The second clique cut separator compared in this experiment, named here as \emph{glpclq}, is provided by the open-source solver of the GNU Linear Programming Kit (GLPK)\footnote{\url{https://www.gnu.org/software/glpk/}} version $4.65$. We ran GLPK with all presolving, preprocessing, heuristics and other cut separators turned off. Thus, we capture only the effect of the inclusion of the clique inequalities. Furthermore, we implemented a callback procedure that computes and stores the number of cuts separated, the current objective value of the LP relaxation, the gap closed and the time elapsed at each iteration of the cut generation loop.

The last clique cut separator that we compared was the one included in the commercial solver IBM ILOG CPLEX\footnote{\url{https://www.ibm.com/analytics/cplex-optimizer}} version 12.8, referred to here as \emph{cpxclq}. We considered only the ``very aggressively'' strategy of this cut separator, since in preliminary experiments it performed slightly better than the other strategies. We ran CPLEX with all presolving, preprocessing, heuristics and other cut separators turned off, according to the CPLEX User’s Manual\footnote{\url{https://www.ibm.com/support/knowledgecenter/SSSA5P_12.8.0/}}. We also implemented a callback procedure that computes and stores the current objective value of the LP relaxation, the gap closed and the time elapsed at each iteration of the cut generation loop.

All cut separators were executed at the root node LP relaxation of the instance problems, considering at most $50$ iterations of the cut generation loop and a time limit of $10{,}800$ seconds. CLP was employed to solve the LP relaxation at each iteration of \emph{bkclq}, \emph{bkclqext} and \emph{cglclq}, while \emph{glpclq} and \emph{cpxclq} used their own linear program solvers.

\begin{figure}[htbp]
\centering
\begin{subfigure}{.45\textwidth}
  \centering
  \includegraphics[width=1.0\linewidth]{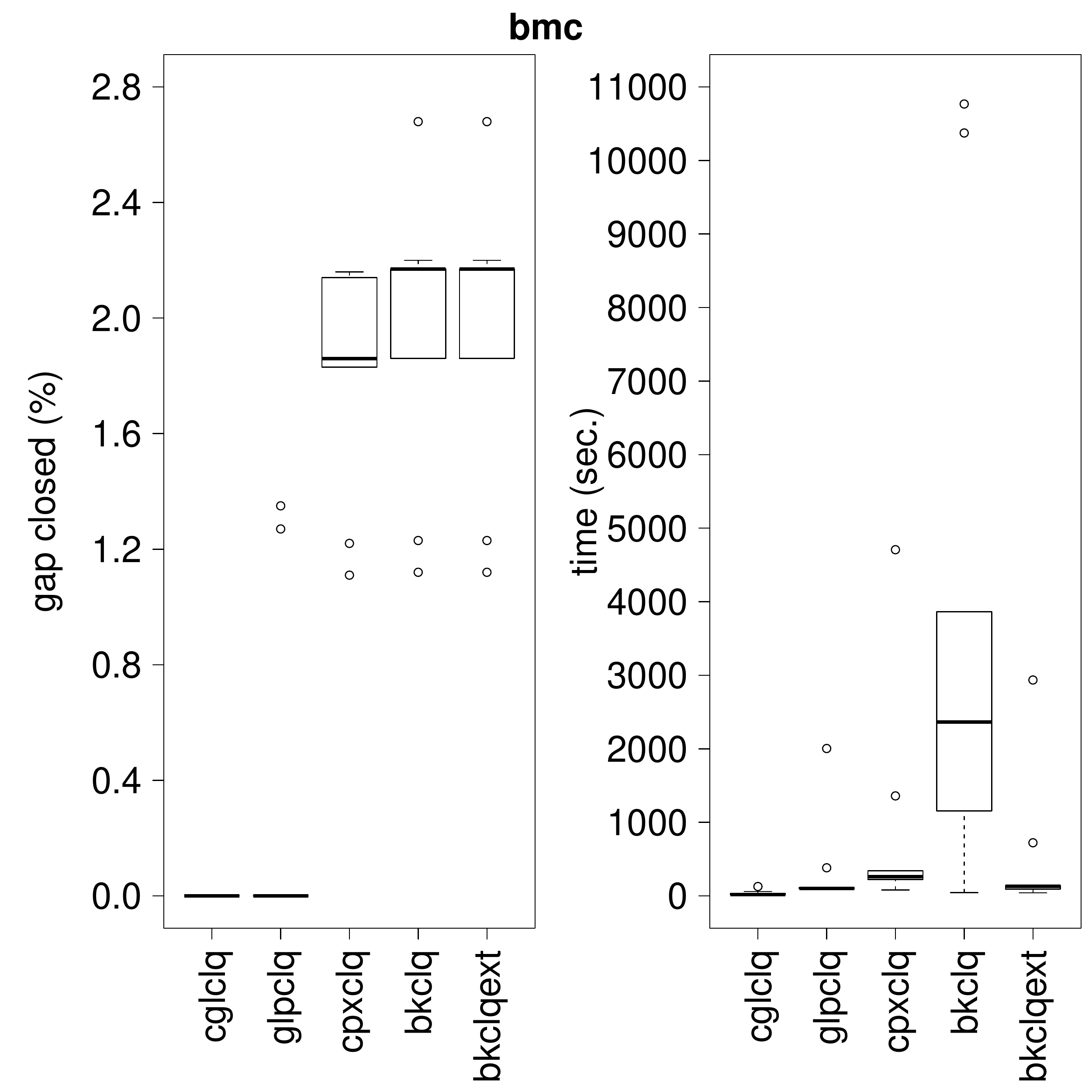}
\end{subfigure}
\begin{subfigure}{.45\textwidth}
  \centering
  \includegraphics[width=1.0\linewidth]{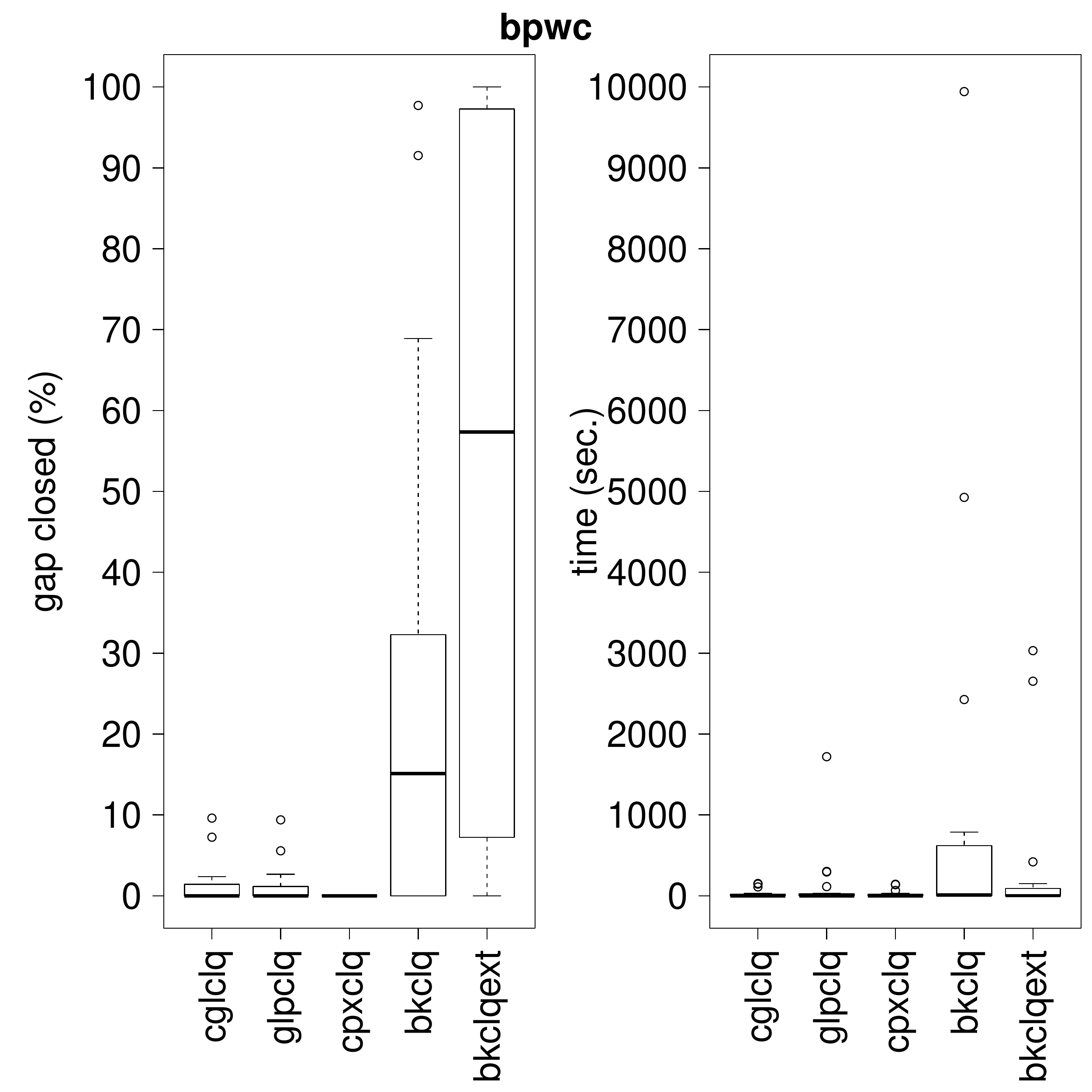}
\end{subfigure}
\begin{subfigure}{.45\textwidth}
  \centering
  \includegraphics[width=1.0\linewidth]{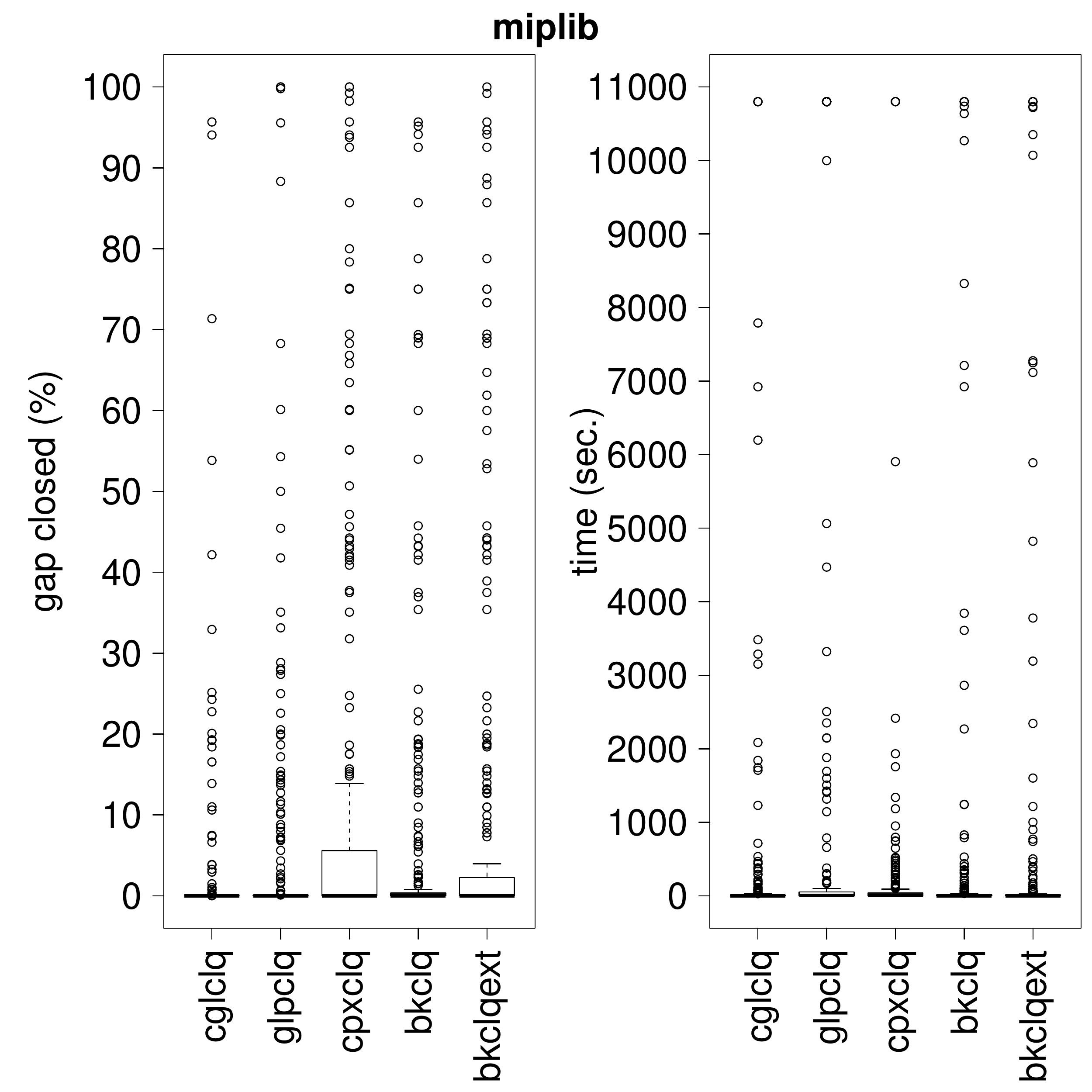}
\end{subfigure}
\begin{subfigure}{.45\textwidth}
  \centering
  \includegraphics[width=1.0\linewidth]{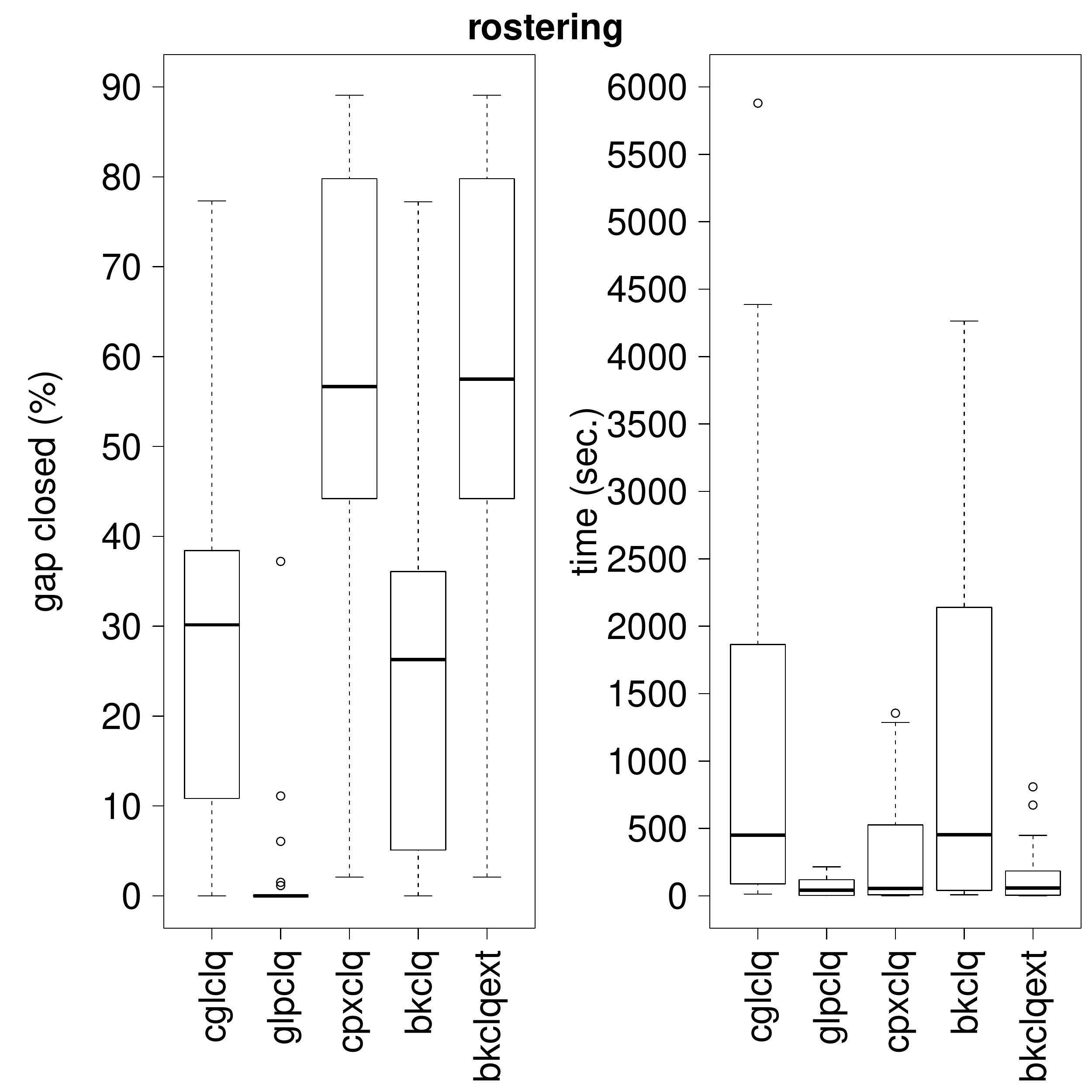}
\end{subfigure}
\begin{subfigure}{.45\textwidth}
  \centering
  \includegraphics[width=1.0\linewidth]{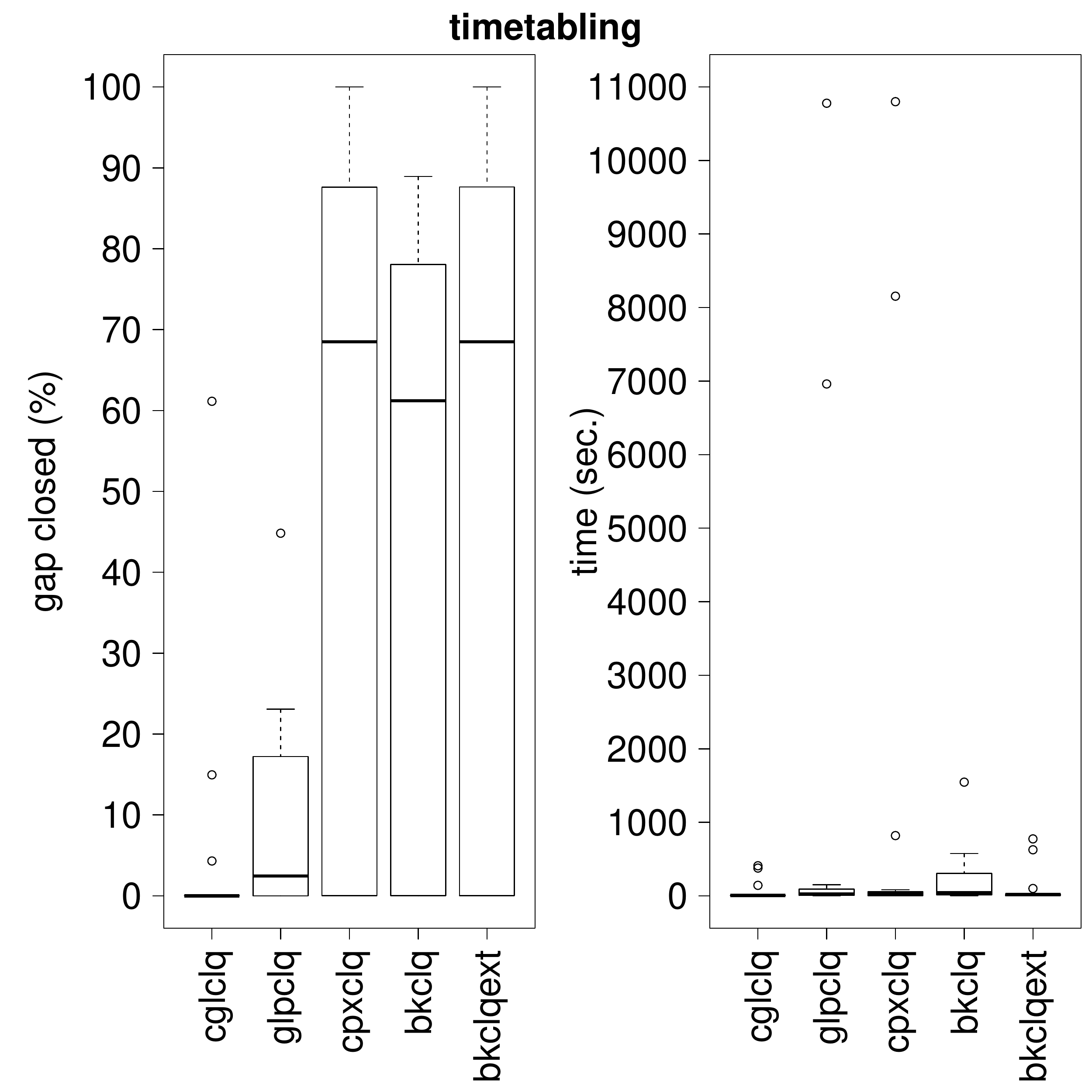}
\end{subfigure}
\caption{Execution times and gap closed by the clique cut separators.}\label{figClqSep}
\end{figure}

Figure~\ref{figClqSep} presents the execution times and gap closed by each clique cut separator. Comparing the two versions of our cut separator, one can observe that the inclusion of the lifting module contributed to improve the execution times. The insertion of lifted cliques avoided some reoptimizations of the LP relaxations, which saved some iterations of the cut generation loop and, consequently, the execution time of the cut separator. Moreover, the version of our clique cut separator that includes the lifting module produced better dual bounds.

Our clique cut separator improved the LP relaxation for $136$ instances. Most instances in which the objective value of the LP relaxation was not changed belong to instance set \emph{miplib}. In fact, we analyzed these instances and noted that their associated CGs have only trivial conflicts or a small set of non-trivial conflicts.

The dual bounds obtained by \emph{bkclqext} are significantly better than those attained by \emph{cglclq} and \emph{glpclq} in all instance sets. Even in instances with denser conflict graphs, such as those in sets \emph{bmc} and \emph{bpwc}, \emph{cglclq} and \emph{glpclq} had difficulty finding violated cliques.

In addition to obtain better dual bounds, the execution of \emph{bkclqext} was responsible for completely closing the gap for a greater number of instances. Cut separators \emph{bkclqext}, \emph{cpxclq} and \emph{glpclq} completely closed the gap for $8$, $5$ and $2$ instances, respectively. The other two cut separators were not able to completely close the gap for any instance.

Similar results in terms of the percentage of gap closed were obtained by \emph{bkclqext} and \emph{cpxclq}, except in set \emph{bpwc}. On the instances of this set, \emph{cpxclq} did not include any violated clique, while the cliques separated by \emph{bkclqext} contributed to close the gap by up $97.71\%$. It is worth mentioning that, for some instances, the initial bounds of the root node LP relaxations computed by CPLEX were already better than the values calculated by CLP and the linear program solver of GLPK, even with the preprocessing routines disabled.

\begin{figure}[htbp]
\begin{center}
	\includegraphics[width=1.0\linewidth]{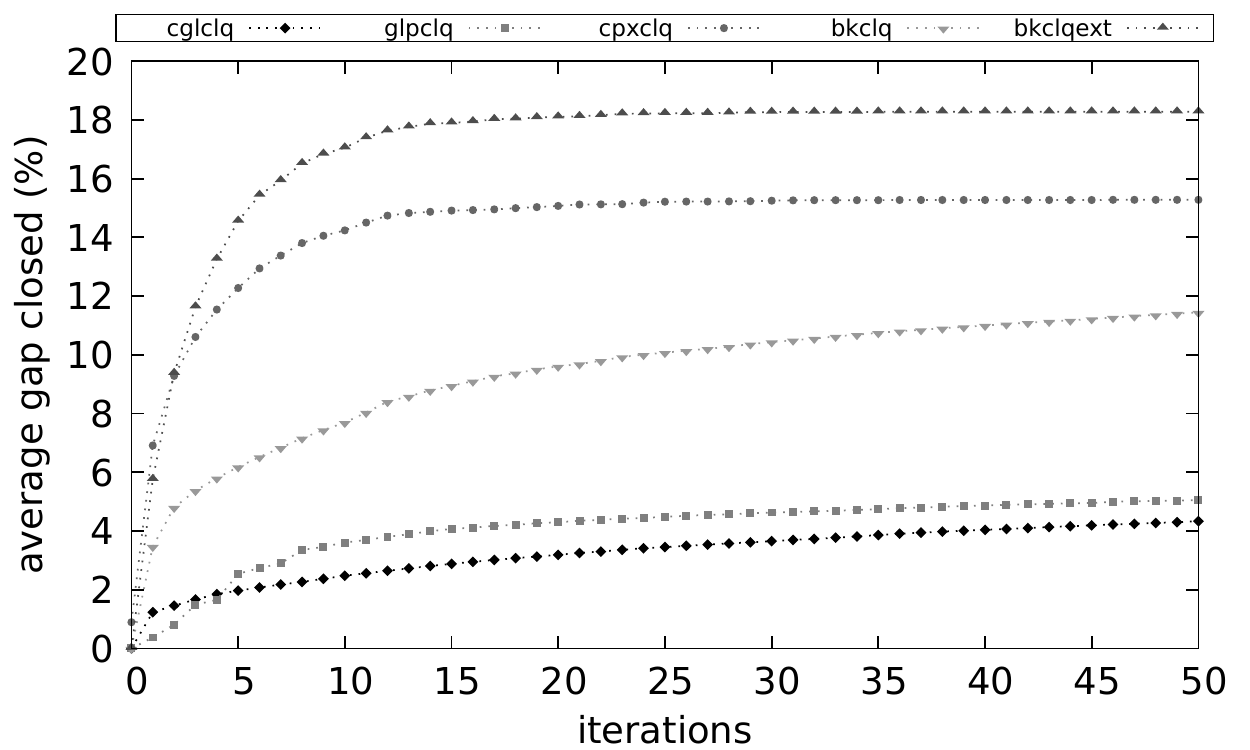}
	\caption{Average gap closed by each clique cut separator.}\label{figCutsAtRoot}
\end{center}
\end{figure}

The average gap closed by the clique cut separators at each iteration of the cut generation loop is presented in Figure~\ref{figCutsAtRoot}. The metric \emph{average gap closed} is also employed in the next experiments and is computed as the arithmetic mean of the gap closed over the problem instances. Differences in the performances of the cut separators can be seen in the earlier steps of the separation process.

After $50$ iterations, the average gap closed by the cut separators was $4.33\%$ for \emph{cglclq}, $5.05\%$ for \emph{glpclq}, $15.27\%$ for \emph{cpxclq} and $18.27\%$ for \emph{bkclqext}. Thus, the average gap closed by our clique cut separator was $4.22$ times better than the clique cut separator of CGL, $3.62$ times better than the clique cut separator provided by GLPK and $19.65\%$ better than the clique cut separator of CPLEX.

\subsection{Odd-Cycle Cut Separator}

Following the evaluation of our conflict graph-based routines, we analyzed the ability of our odd-cycle cut separator in improving the bounds given by the LP relaxation of a MILP. We also investigated the impact of performing our proposed lifting module, which consists of inserting a clique into the center of an odd cycle. For this, we ran our odd-cycle cut separation at the root node LP relaxation of the instances, considering at most $50$ iterations of the cut generation loop and a time limit of $10{,}800$ seconds. CLP was employed to solve the LP relaxation at each iteration. The metric used for comparison was the gap closed, which is calculated by Equation~\ref{gapClosed}.

\begin{figure}[htbp]
\centering
\begin{subfigure}{.45\textwidth}
  \centering
  \includegraphics[width=1.0\linewidth]{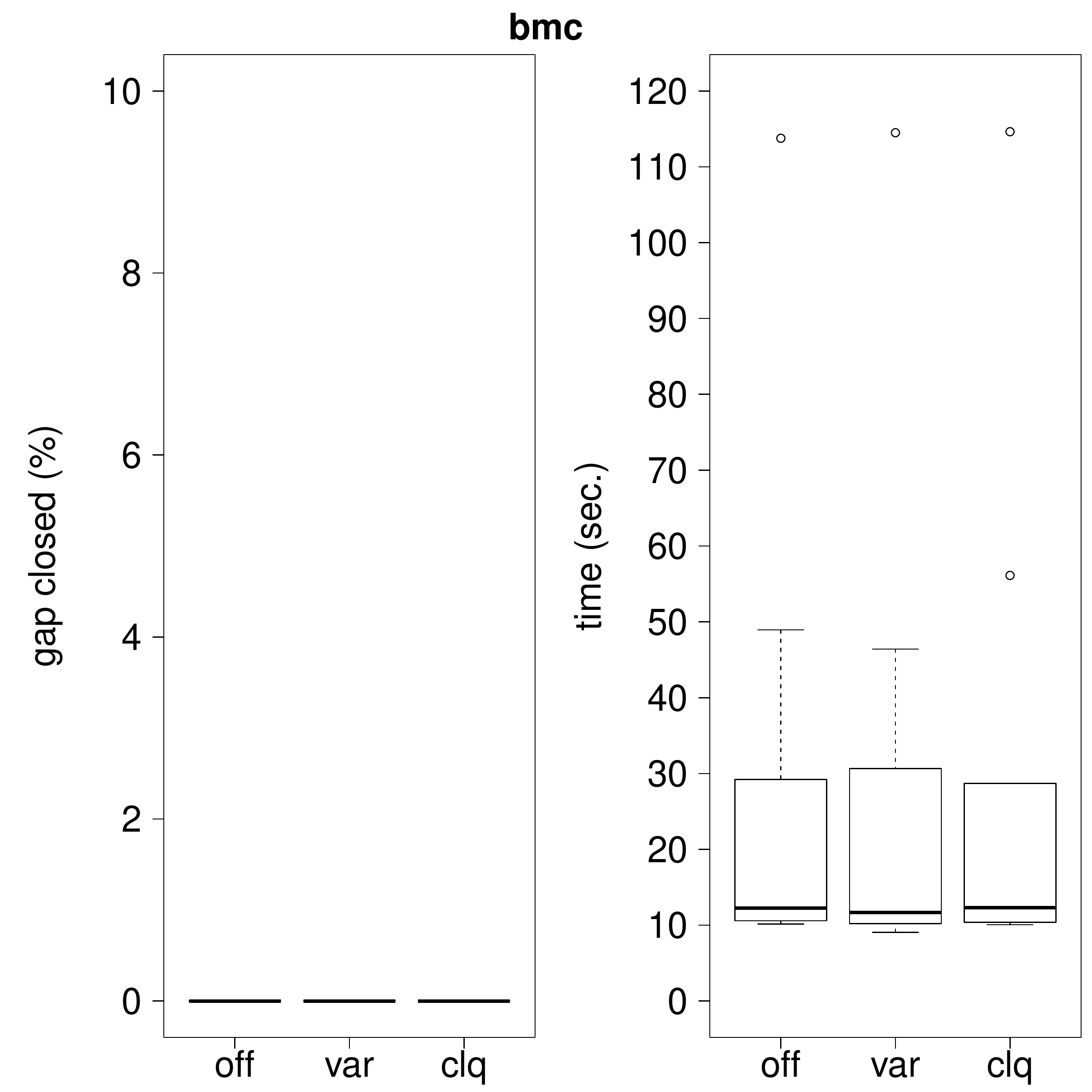}
\end{subfigure}
\begin{subfigure}{.45\textwidth}
  \centering
  \includegraphics[width=1.0\linewidth]{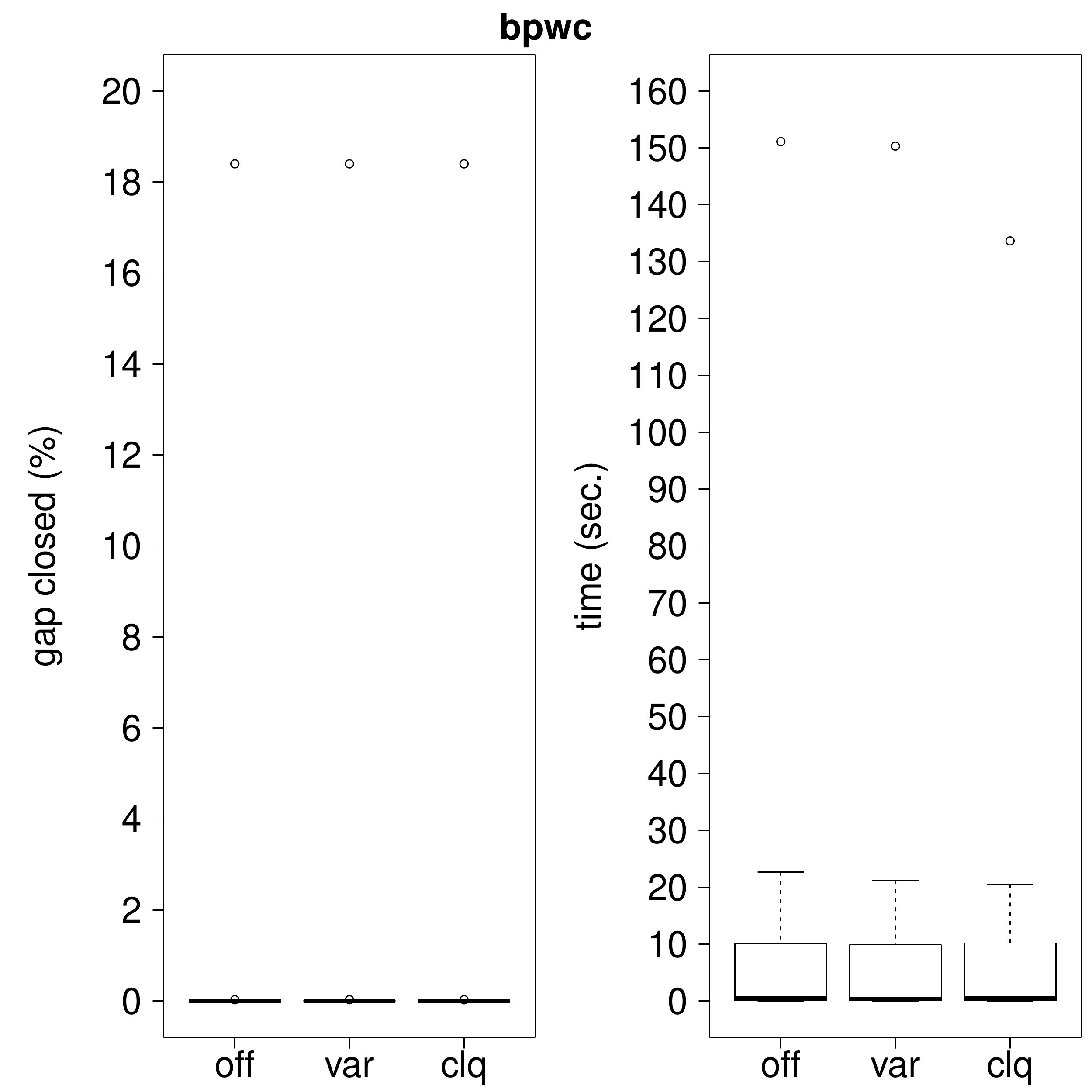}
\end{subfigure}
\par\bigskip
\begin{subfigure}{.45\textwidth}
  \centering
  \includegraphics[width=1.0\linewidth]{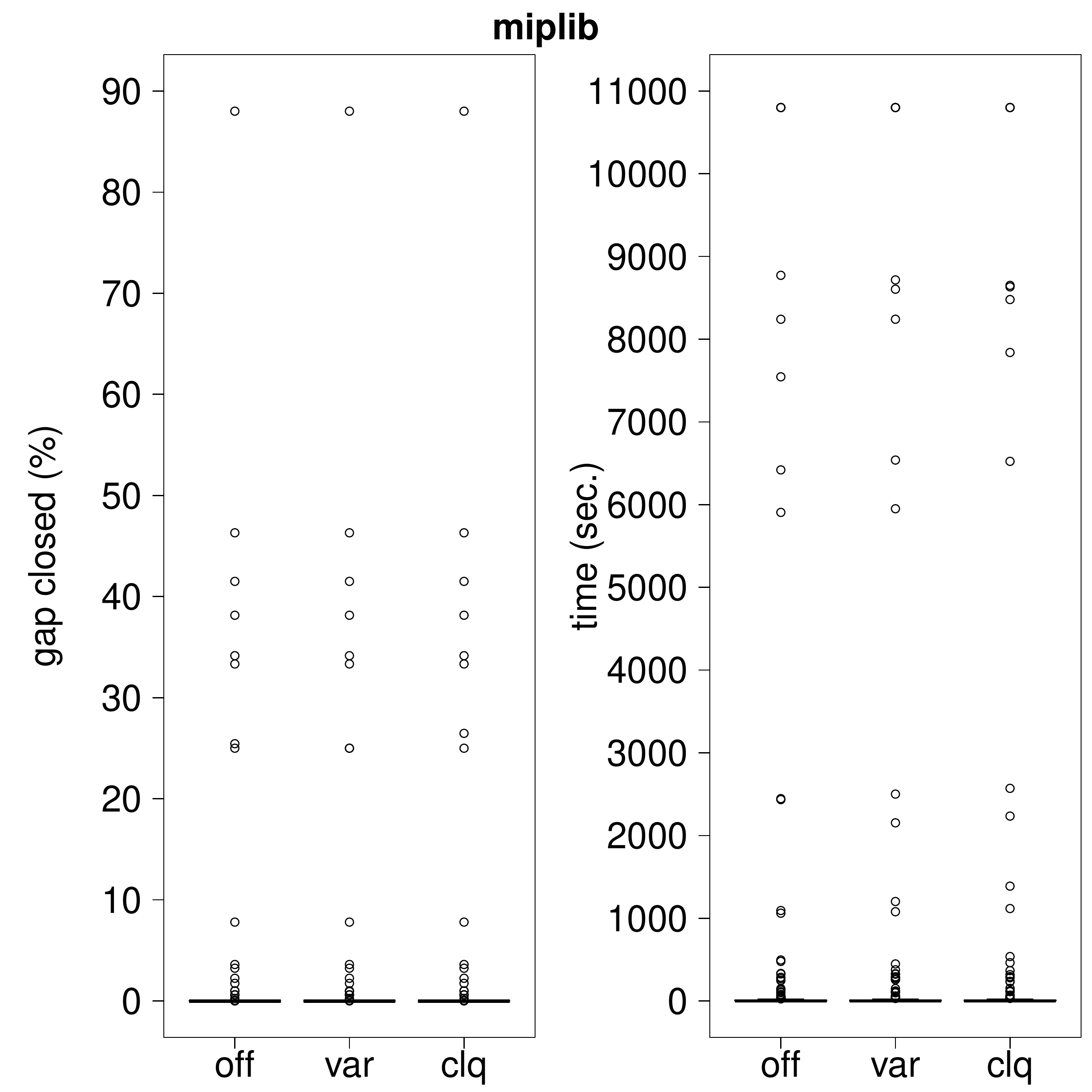}
\end{subfigure}
\begin{subfigure}{.45\textwidth}
  \centering
  \includegraphics[width=1.0\linewidth]{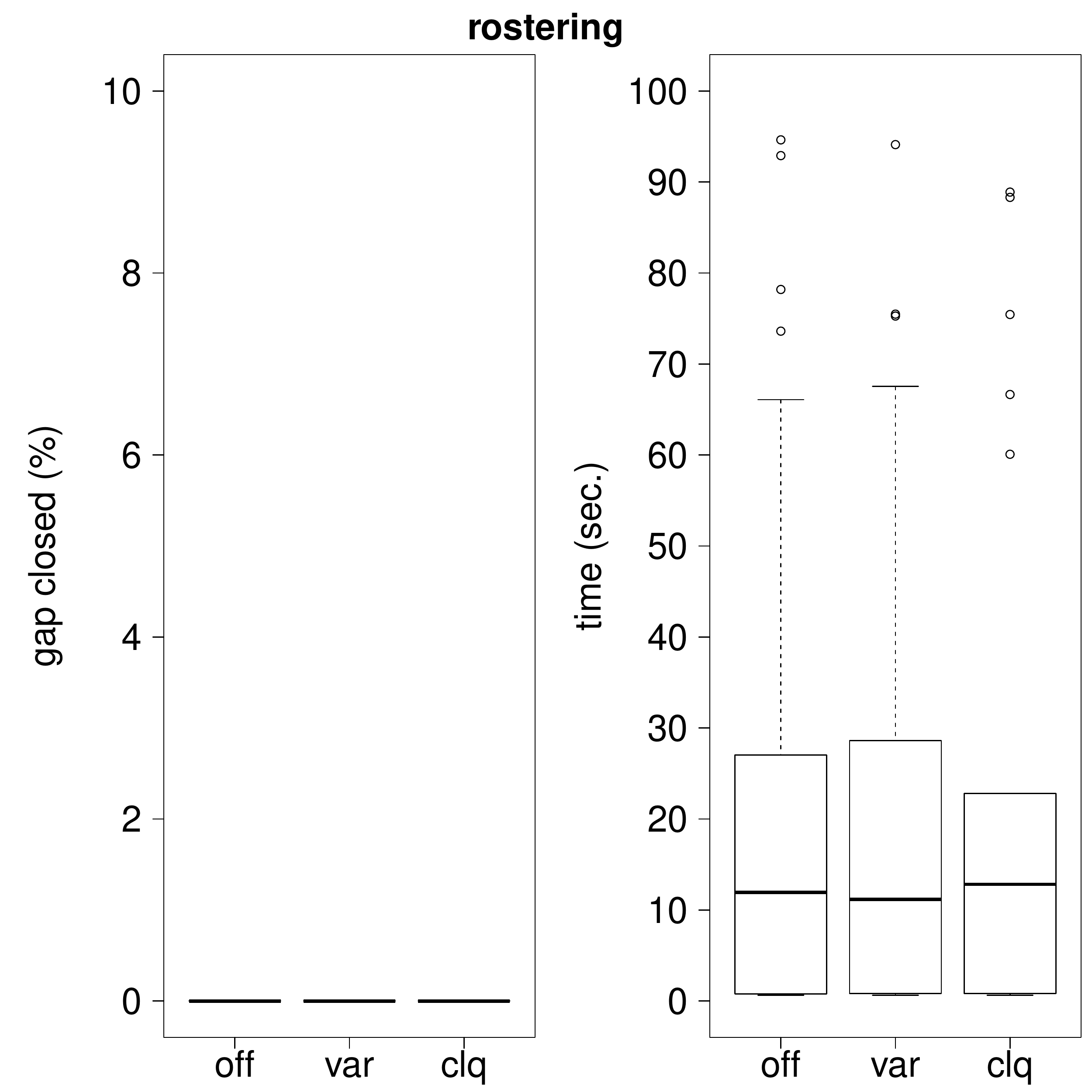}
\end{subfigure}
\par\bigskip
\begin{subfigure}{.45\textwidth}
  \centering
  \includegraphics[width=1.0\linewidth]{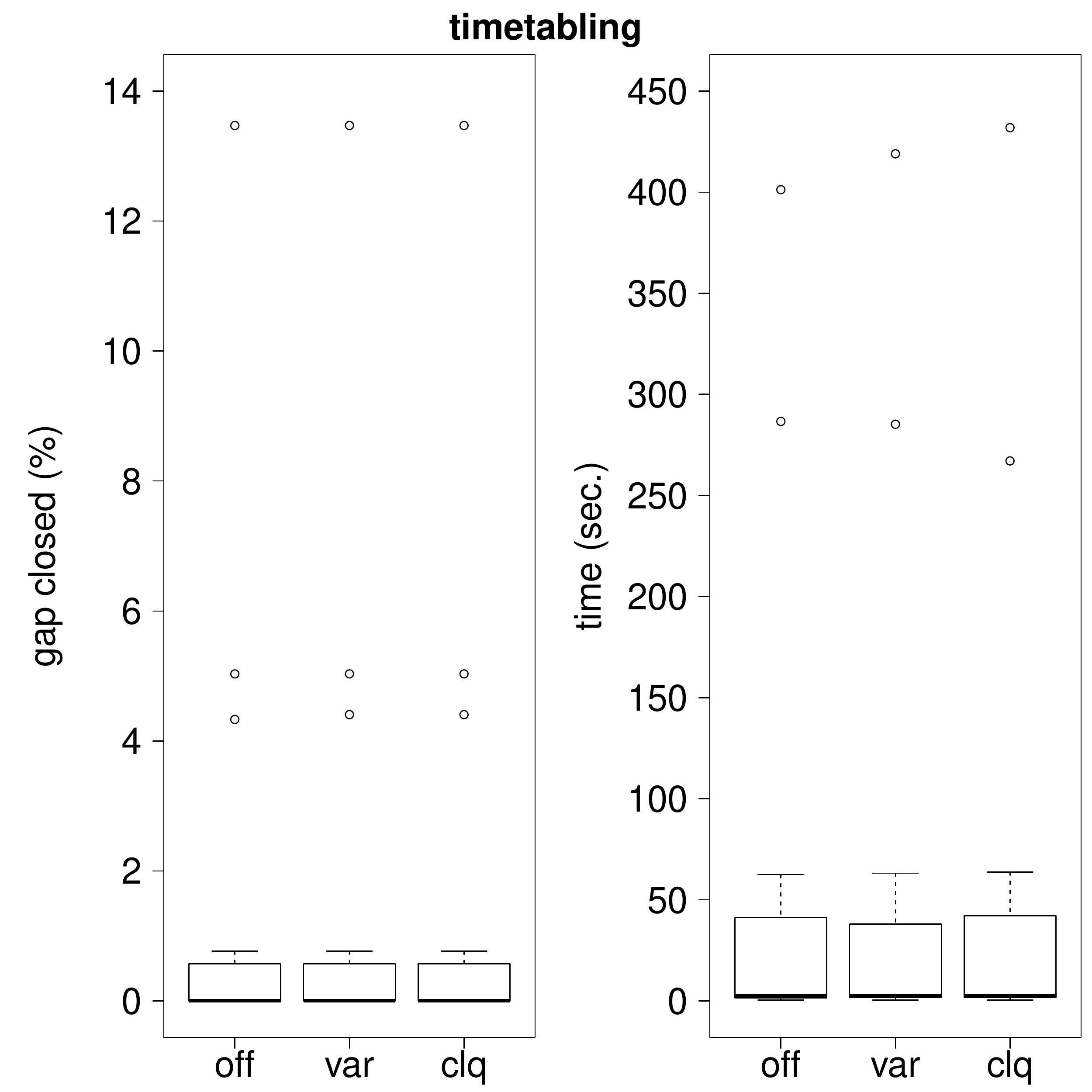}
\end{subfigure}
\caption{Execution times and gap closed by the odd-cycle cut separator.}\label{figOddW}
\end{figure}

We ran three versions of our cut separator. In the first version, named here as \emph{off}, we did not execute the lifting module. In the second version, referred to here as \emph{var}, we executed a lifting module that tries to insert one variable into the center of an odd cycle. Our proposed lifting module was performed in the third version of the cut separator and is referred to here as \emph{clq}. The results of this experiment are presented in Figure~\ref{figOddW}.

Regardless of the version of the cut separator, the inclusion of odd-cycle inequalities had no significant impact on the dual bound improvement. For most instances, no odd cycles of size greater than three were found. As explained before, odd cycles of size three are not separated by the odd-cycle cut separator, since they correspond to cliques and can be separated by the clique cut separator. Even in instances where a considerable set of odd-cycle cuts were separated, the improvement in the LP relaxation was small. For example, \emph{clq} separated $1{,}764$ odd-cycle cuts in instance \emph{br2} of set \emph{timetabling}, but the gap closed was $5.03\%$.

Odd-cycle cuts were separated in $82$ of $320$ instances, but the improvement in the objective value of the LP relaxation occurred only in $31$ of them. The maximum percentage of gap closed by our cut separator was $88.00\%$, obtained by all separators in instance \emph{neos8}. Several odd cycles of size $5$, $7$ and $9$ were separated in this instance.

In general, the time spent in separating odd-cycle cuts was small. Considering all problem instances, the maximum time spent in this step was $7.42$ seconds. The execution of CLP to solve the LP relaxation of the problems in each iteration of the cut generation loop was responsible for increasing the execution times.

It was not possible to detect differences between the performances of the three versions of our odd-cycle separator. In fact, there were few cases where the lifting module was able to insert wheels into the centers of the odd cycles. Regardless of the version of the lifting module, some odd cycles were transformed into odd wheels only in $11$ instances. For these instances, the average gap closed by \emph{clq} was $3.84\%$ better than \emph{var}. Only in $5$ instances a clique was added instead of just a vertex as the center of a wheel.

\subsection{Improving COIN-OR Branch-and-Cut Solver}

Previous experiments were performed to evaluate the performance of our routines for building CGs, preprocessing MILPs and separating cuts. After that, we generated a new version of the CBC solver, including our algorithms and data structures in the source code of its development version. In this version, a conflict graph is constructed after the execution of the preprocessing routines of CBC, followed by the execution of the clique strengthening routine. Our conflict-based cut separators are performed during the execution of the branch-and-cut algorithm. The clique and odd-cycle cut separators of CGL, included in the previous version of CBC, were replaced by our cut separators. The new version of CBC are denoted here as \emph{cbc+cg}.

\begin{figure}[htbp]
\centering
\begin{subfigure}{.45\textwidth}
  \centering
  \includegraphics[width=1.0\linewidth]{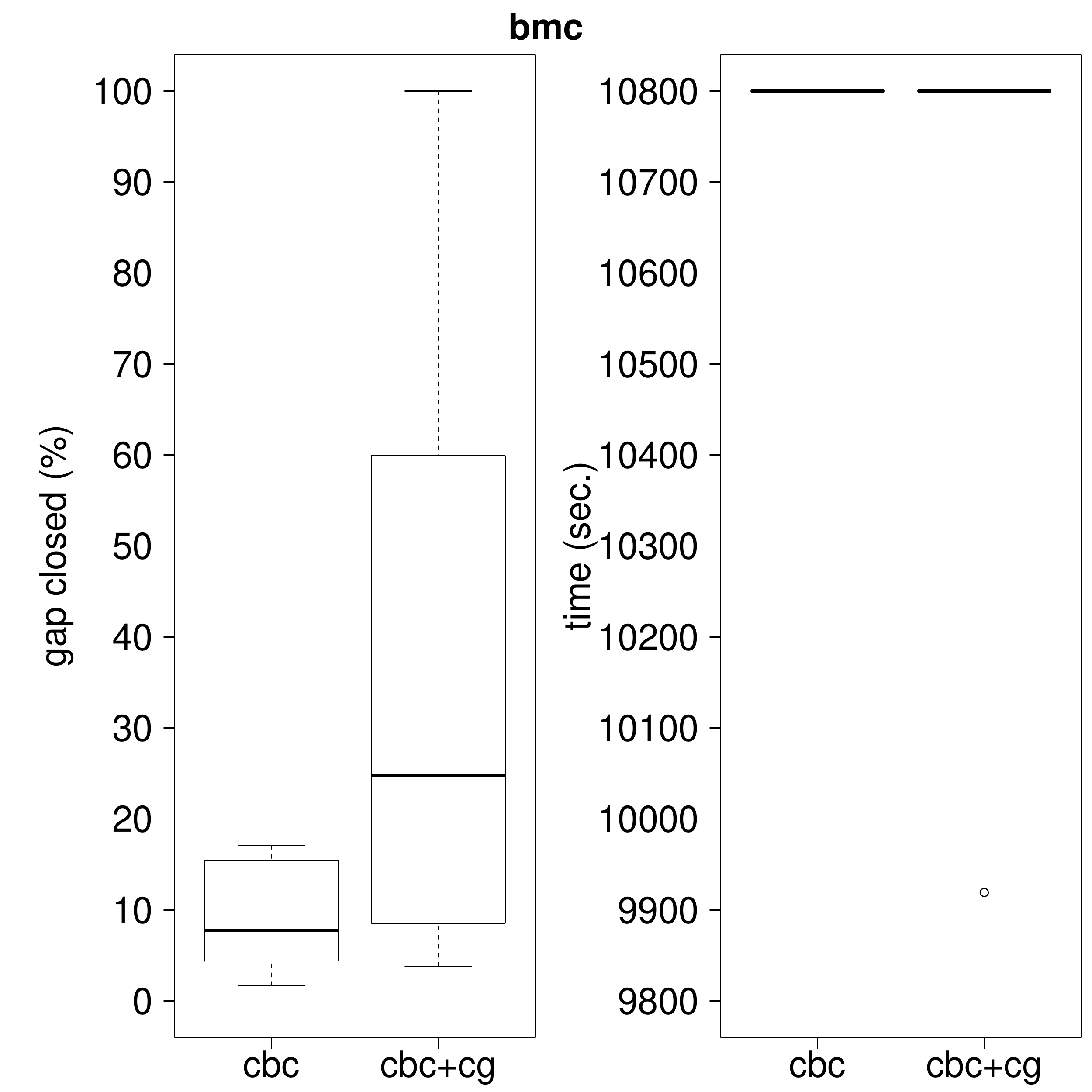}
\end{subfigure}
\begin{subfigure}{.45\textwidth}
  \centering
  \includegraphics[width=1.0\linewidth]{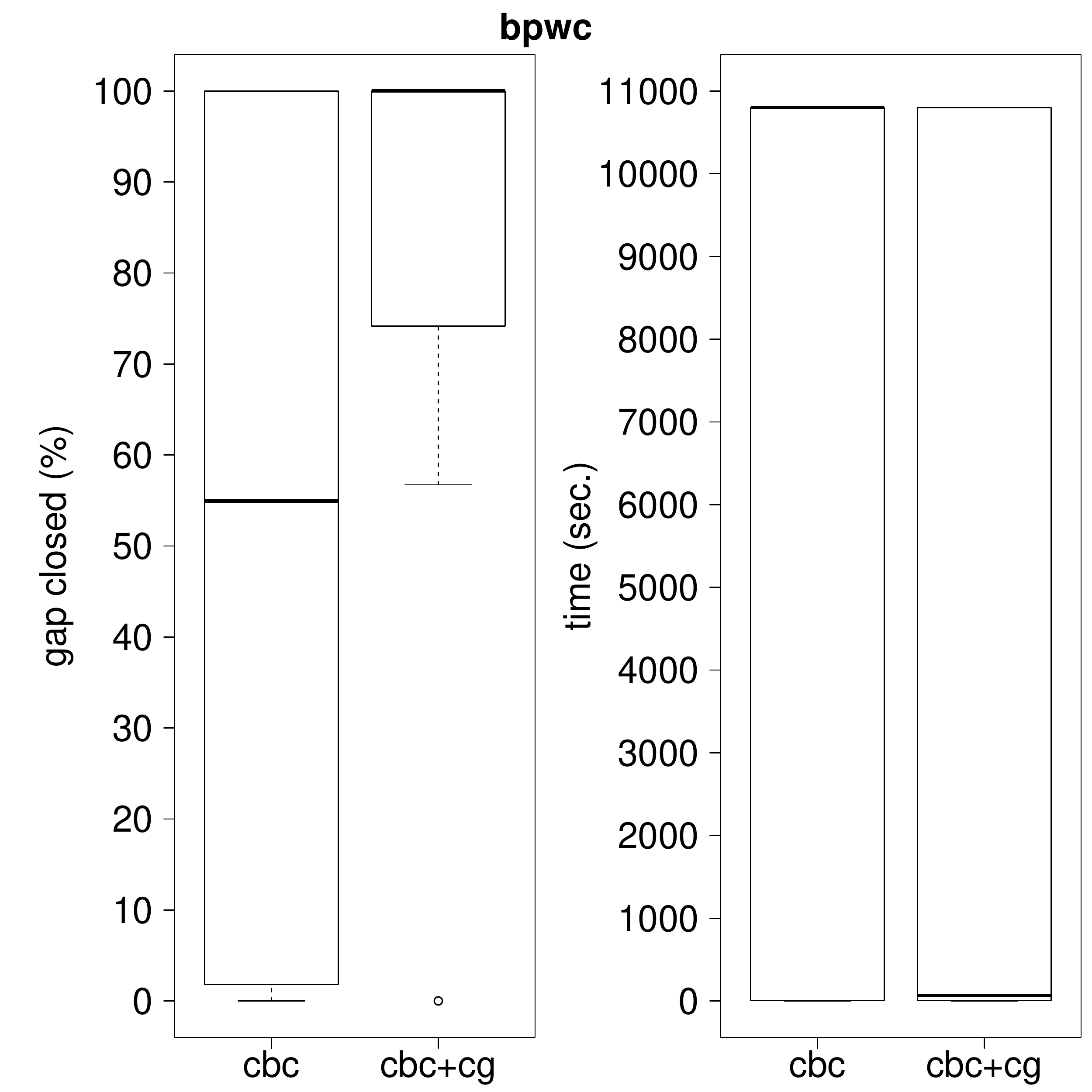}
\end{subfigure}
\par\bigskip
\begin{subfigure}{.45\textwidth}
  \centering
  \includegraphics[width=1.0\linewidth]{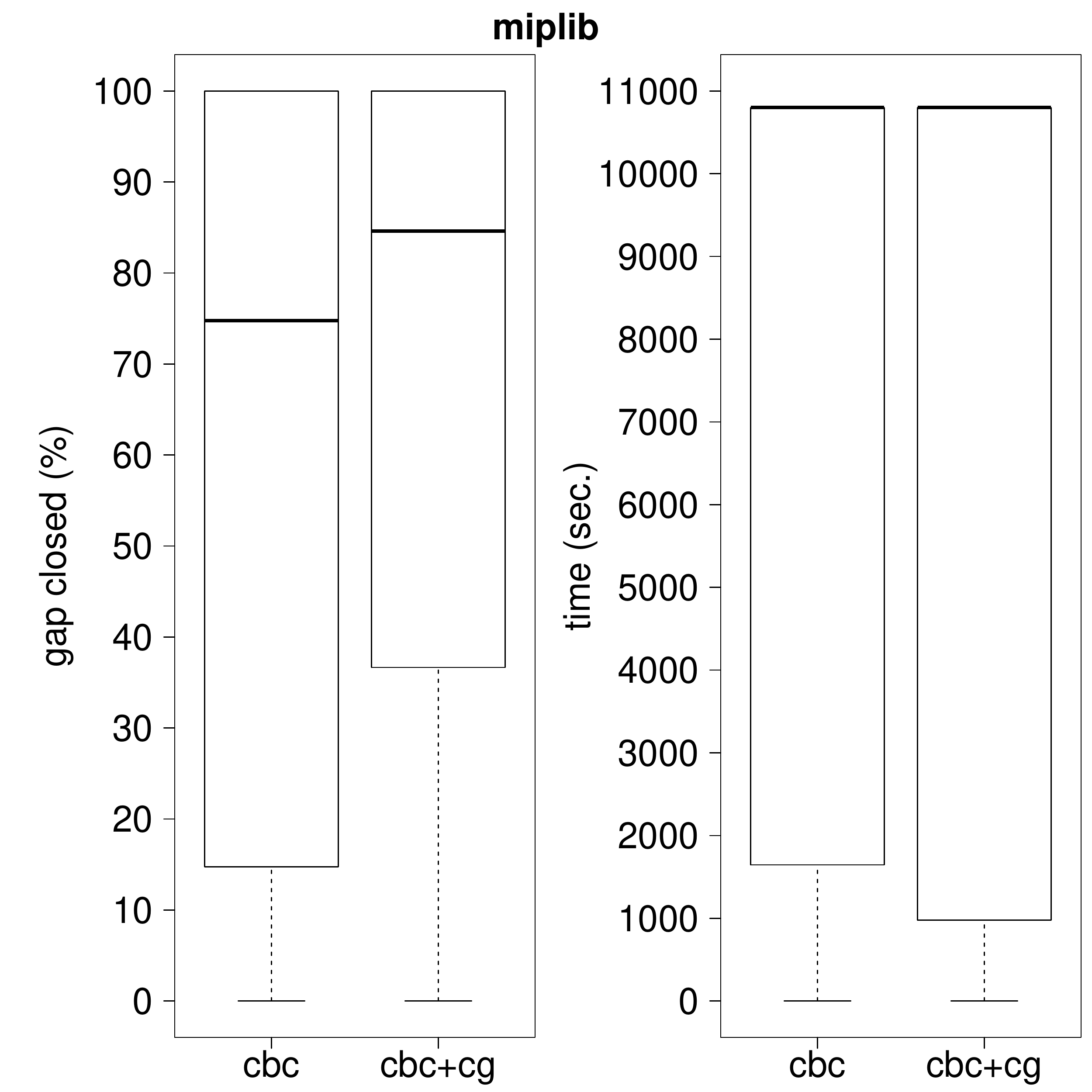}
\end{subfigure}
\begin{subfigure}{.45\textwidth}
  \centering
  \includegraphics[width=1.0\linewidth]{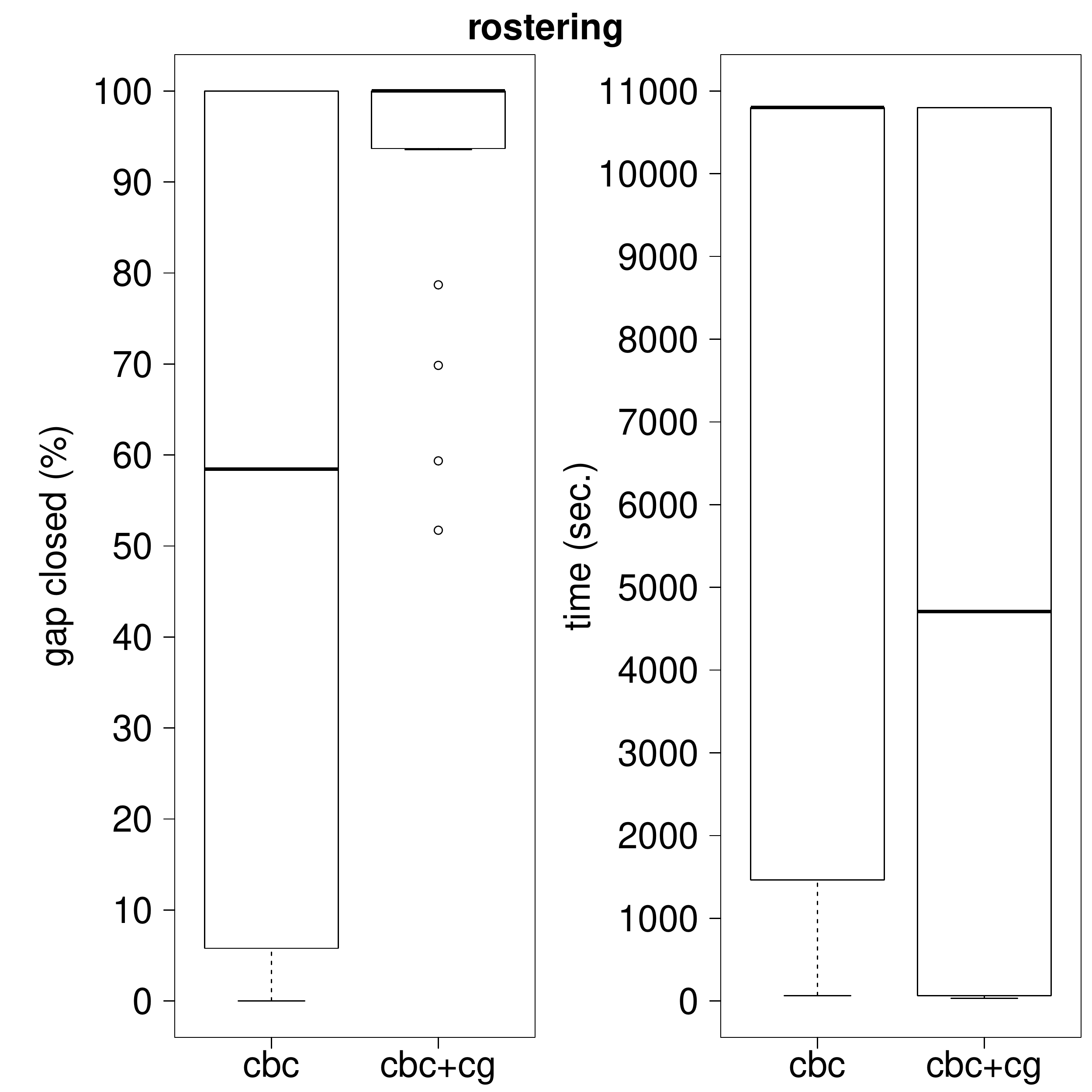}
\end{subfigure}
\par\bigskip
\begin{subfigure}{.45\textwidth}
  \centering
  \includegraphics[width=1.0\linewidth]{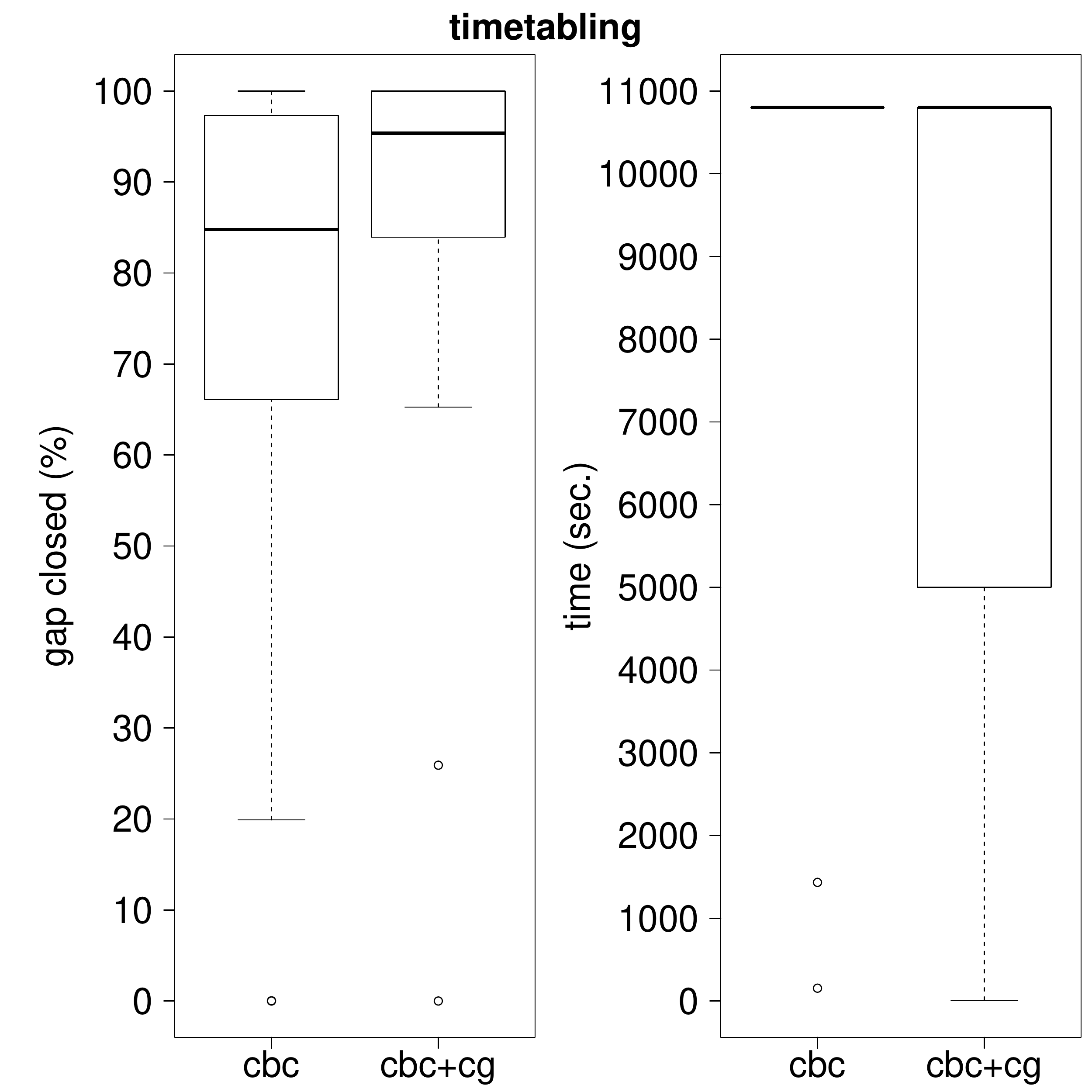}
\end{subfigure}
\caption{Execution times and gap closed by the two versions of CBC solver.}\label{figBB}
\end{figure}

As the last experiment, we investigated the performance improvement of the new version of CBC against its previous version, referred to here as \emph{cbc}. The default parameters of CBC for heuristics, preprocessing, branching rules and cuts separators were used in both versions. We ran each version on the $320$ MILPs of our dataset considering a time limit of $10{,}800$ seconds. Figure~\ref{figBB} shows the results of this experiment.

The inclusion of our conflict graph-based algorithms and data structures in CBC contributed significantly to improve the dual bounds obtained by this solver. As observed in Figure~\ref{figBB}, the median gap closed by the new version of CBC is greater than the values obtained by the previous version of this solver in all instance sets. The percentage of gap closed by \emph{cbc+cg} was greater than or equal to that obtained by \emph{cbc} in all $320$ instances. Consequently, the average gap closed by CBC increased from $58.86\%$ to $68.76\%$, representing an improvement of $16.82\%$. Reductions in the execution times were observed in several instances.

\begin{figure}[htbp]
\centering
\begin{subfigure}{.45\textwidth}
  \centering
  \includegraphics[width=1.0\linewidth]{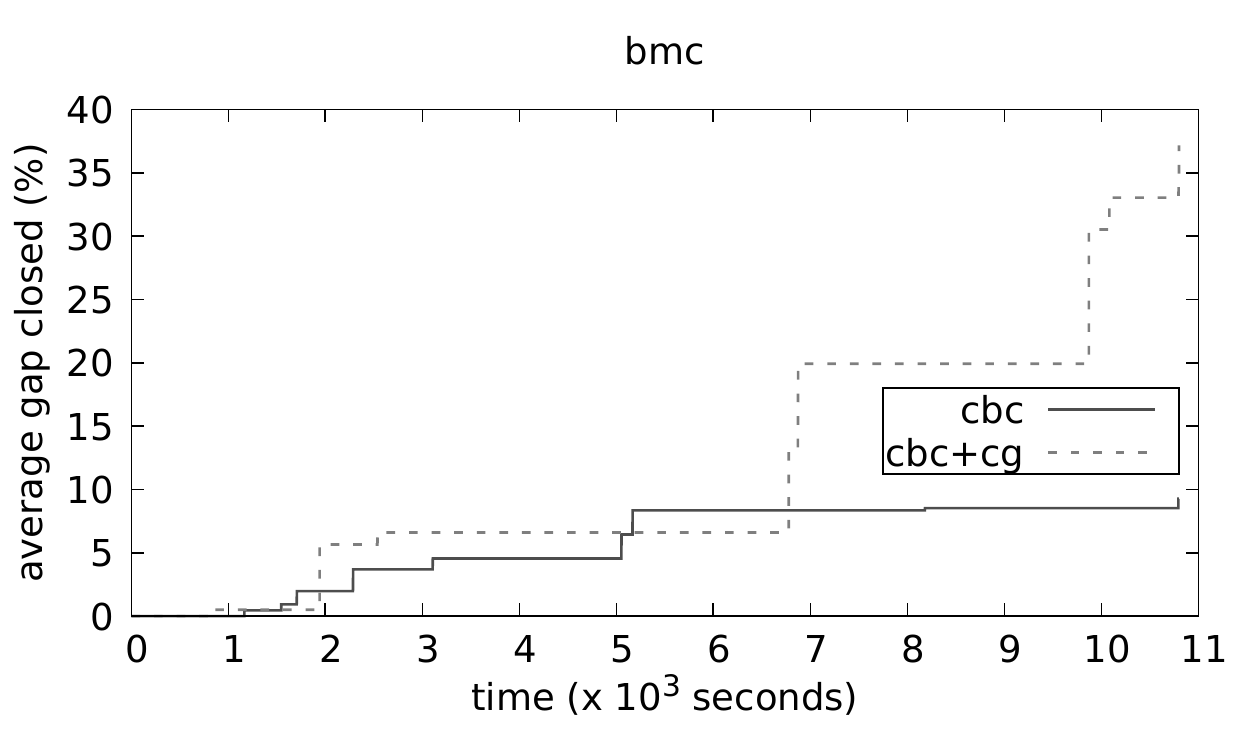}
\end{subfigure}
\begin{subfigure}{.45\textwidth}
  \centering
  \includegraphics[width=1.0\linewidth]{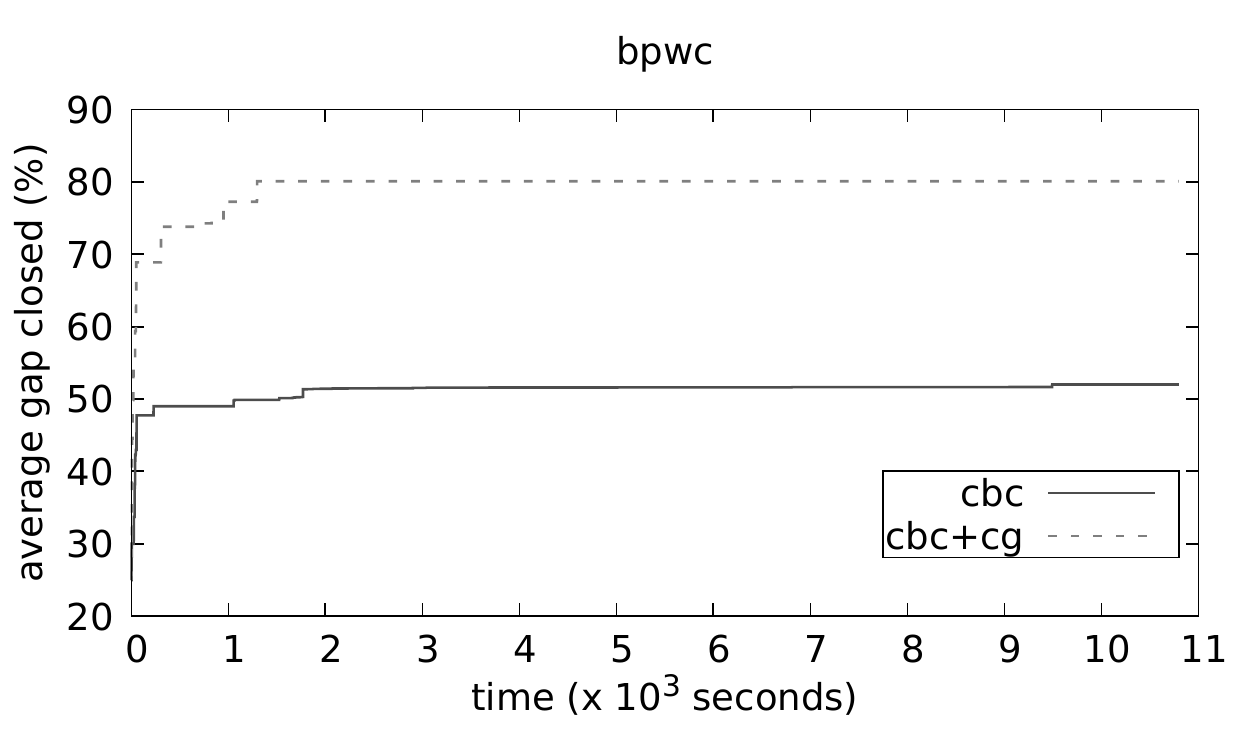}
\end{subfigure}
\begin{subfigure}{.45\textwidth}
  \centering
  \includegraphics[width=1.0\linewidth]{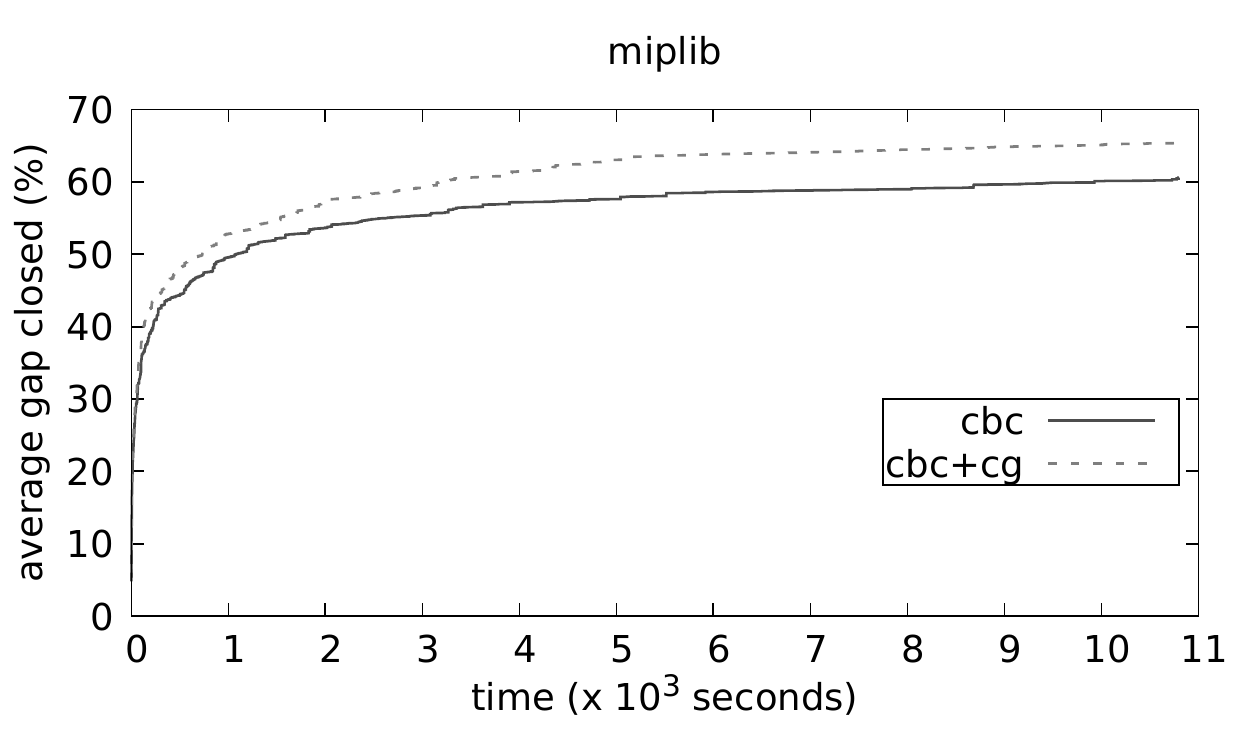}
\end{subfigure}%
\begin{subfigure}{.45\textwidth}
  \centering
  \includegraphics[width=1.0\linewidth]{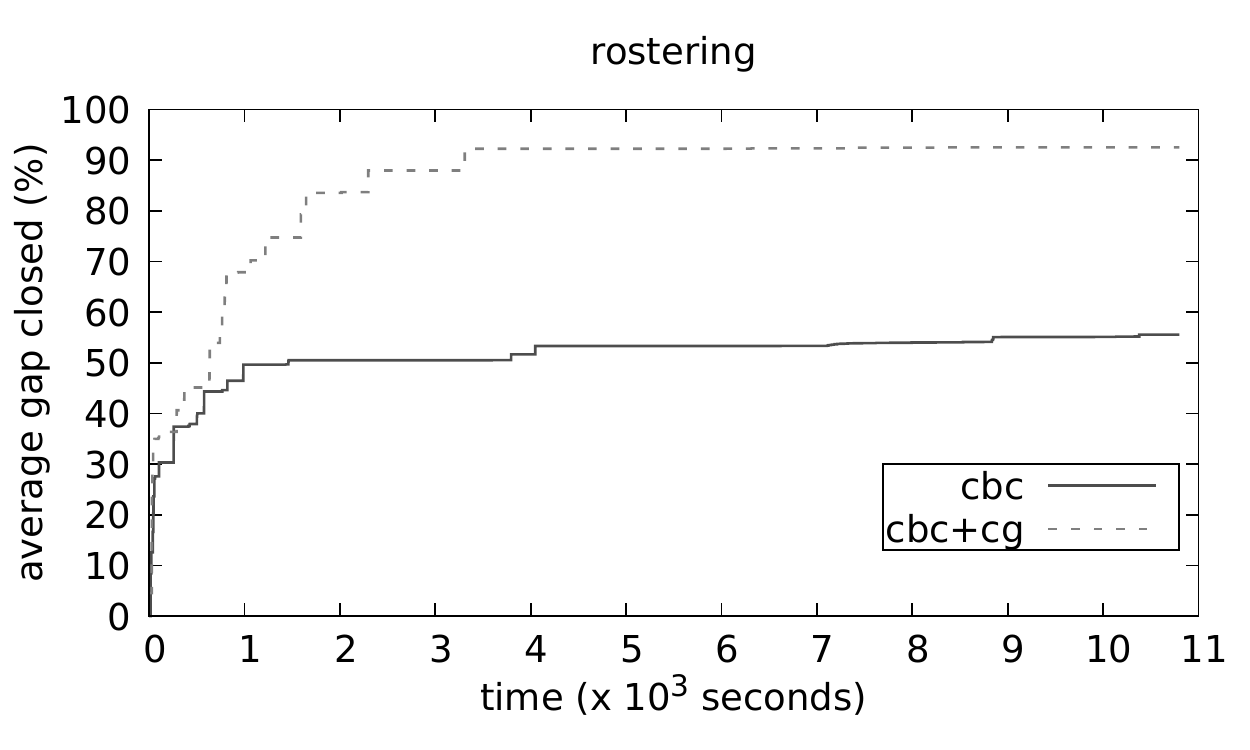}
\end{subfigure}
\begin{subfigure}{.45\textwidth}
  \centering
  \includegraphics[width=1.0\linewidth]{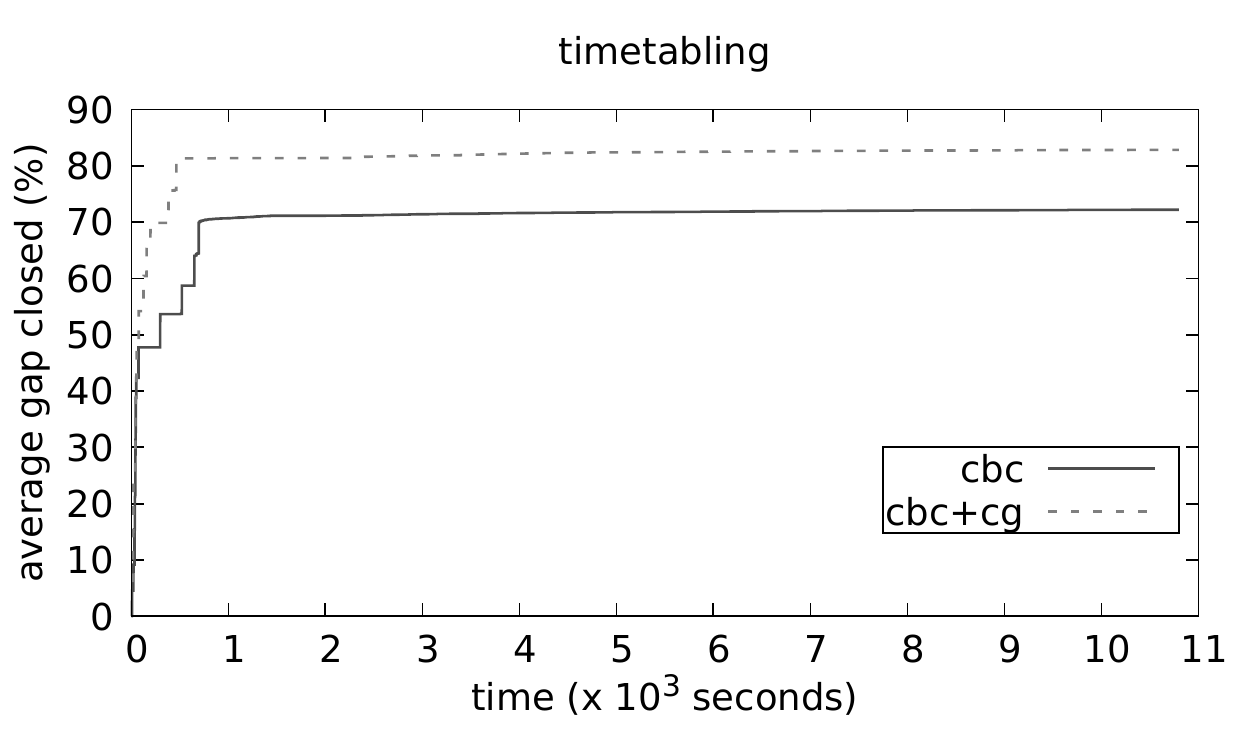}
\end{subfigure}%
\caption{Evolution of the average gap closed over time for each instance set.}\label{figBBGapGroups}
\end{figure}

In order to better visualize the results, we computed the evolution of the average gap closed by each version of CBC over the execution time for each instance set. The results are presented in Figure~\ref{figBBGapGroups}. The most significative improvements were obtained in instance sets \emph{bmc}, \emph{bpwc} and \emph{rostering}. In instances of these three sets, the clique strengthening routine considerably reduced the number of rows of these instances, while the clique cut separator inserted strong valid inequalities. 

At the end of the executions, the average gap closed by \emph{cbc+cg} was four times better than the one obtained by \emph{cbc} in instance set \emph{bmc}. Furthermore, the average gap closed by the new version of CBC was $54.01\%$ better in instance set \emph{bpwc}, and $66.45\%$ better in \emph{rostering}.

Improvements in the average gap closed of instance sets \emph{miplib} and \emph{timetabling} were slightly smaller. In \emph{timetabling}, the average gap closed by CBC increased in $14.74\%$. The smallest improvement in the average gap closed was $8.95\%$, obtained in the instances of \emph{miplib}. As observed in the previous experiments, the conflict graphs of several instances of this set have only trivial conflicts or a small set of non-trivial conflicts. Thus, conflict-based routines have difficulties to improve the dual bounds.

\begin{figure}[htbp]
\begin{center}
	\includegraphics[width=1.0\linewidth]{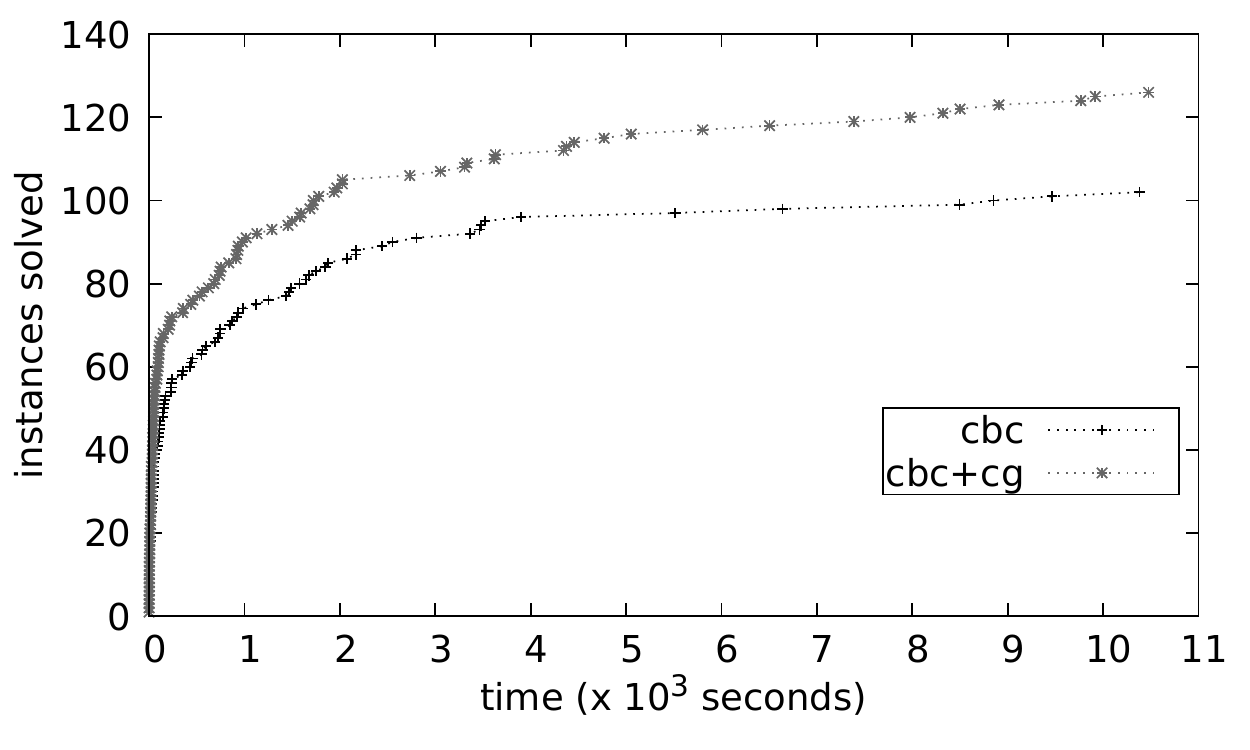}
	\caption{Number of instances solved over three hours.}\label{figBBOptimal}
\end{center}
\end{figure}

We also investigated the evolution of the number of instances solved by each version of CBC over time. These results are provided in Figure~\ref{figBBOptimal}.

The previous version of CBC was able to prove the optimality for $102$ instances, while the new version proved the optimality for $126$ instances. This result represents an increase of $23.53\%$ in the number of instances solved. Moreover, the use of our conflict graph-based algorithms and data structures not only increased the number of instances solved but also decreased the execution time necessary for doing so. Almost half of $24$ instances in which \emph{cbc+cg} proven the optimality and \emph{cbc} stopped by time limit were solved in less than $1{,}000$ seconds.

\section{Conclusions and Future Work}\label{conclusions}

This paper presented conflict graph-based algorithms and data structures for Mixed-Integer Linear Programming problems. We developed a conflict graph infrastructure, characterized by the efficient construction and storage of such graphs. Our algorithm for building conflict graphs is an improved version of a state-of-the-art conflict detection algorithm that extracts cliques from the MILP constraints. An additional step that detects additional maximal cliques was included in this algorithm, without changing its worst-case complexity. We also developed optimized data structures that selectively store conflicts pairwise or grouped in cliques and exploit the sequence in which similar cliques are discovered to store them compactly. Our conflict graph infrastructure was able to construct graphs even for instances with a large number of conflicts.

Conflict graphs were then used in the implementation of a preprocessing algorithm and two cut separators. The preprocessing algorithm is based on a clique strengthening procedure that combines several set packing constraints into a single constraint. Significant improvements with respect to the dual bounds were obtained, especially for MILPs with several constraints including a relatively small number of conflicting variables. Our preprocessing routine was responsible for reducing the number of constraints, strengthening the initial dual bounds and for accelerating the process of proving optimality for a great number of instances.

The two conflict-based cut separators that we developed are responsible for separating cliques and odd cycles. Our clique cut separator obtained better dual bounds than those provided by the equivalent cut generators of CBC, GLPK and CPLEX solvers. As previous works show, the inclusion of odd-cycle cuts had no significant impact on the dual bound improvement. However, the cost for separating these cuts is low, which means that they can be included in a cutting plane strategy without a significant increase in execution time.

A new version of CBC solver was generated which the inclusion of our conflict graph infrastructure, preprocessing routine and cut separators. Experiments with this new version of CBC revealed an improvement in the average gap closed of $16.82\%$ in comparison to the previous version of this solver. For some instance sets, the average gap closed by CBC was improved up to four times. Furthermore, the overall time spent to prove optimality for the instances decreased, while the number of instances solved increased from $102$ to $126$.

As future work, we intend to develop a conflict-based heuristic for generating initial feasible solutions for MILPs. Machine learning techniques could be used to decide when to activate or deactivate the preprocessing and cut separators since for some cases they cannot improve dual bounds. Finally, the use of constraint programming techniques for detecting additional conflicts in the graphs appears to be a promising future research direction.

\section*{Acknowledgements}
We acknowledge UFOP, CNPq and FAPEMIG for supporting the development of this research and Luke Connolly (KU Leuven) for providing editorial consultation on the preliminary version of this paper. We would also like to thank the Combinatorial Optimization and Decision Support (CODeS) research group in KU Leuven by providing the senior research fellowship in 2018-2019 to Prof. Haroldo G. Santos.

\bibliography{references}

\end{document}